\documentclass[twocolumn,aps,prb,10pt]{revtex4-1} 
 
\usepackage{graphicx} 
\usepackage{times} 
\usepackage[utf8]{inputenc} 
\usepackage{amsmath}
\usepackage{amsfonts}

\begin{document}

\title{Optical and Electrical Properties of Nanostructured Metallic
  Electrical Contacts}
\author{Victor J. Toranzos}
\author{Guillermo P. Ortiz}
\affiliation{{\em Departamento de F\'{\i}sica - Facultad de Ciencias
    Exactas Naturales y Agrimensura, Universidad Nacional del
    Nordeste, Corrientes, Argentina.\\gortiz@exa.unne.edu.ar\\}} 
\author{W. Luis Moch\'an}
\affiliation{Instituto de Ciencias F\'{\i}sicas, Universidad Nacional Aut\'onoma de M\'exico, \\Cuernavaca, Morelos, M\'exico}
\author{Jorge O. Zerbino}
\affiliation{{\em Instituto de Investigaciones Fisicoqu\'{\i}micas Te\'oricas y Aplicadas\\Centro Investigaciones Cient\'{\i}ficas de la Provincia de Buenos Aires, La Plata, Argentina.}}

\begin{abstract} 
We study the optical and electrical properties of silver films with a
graded thickness obtained through metallic evaporation in
vacuum on a tilted substrate to evaluate their use as semitransparent
electrical contacts. We measure their ellipsometric coefficients,
optical transmissions and electrical conductivity for different
widths, and we employ
an efficient recursive method to calculate their macroscopic
dielectric function, 
their optical properties and their microscopic electric fields. The
topology of very thin films corresponds to disconnected islands, while
very wide films are simply connected. For intermediate widths the film
becomes semicontinuous, multiply connected, and its
microscopic electric field develops hotspots at 
optical resonances which appear near
the percolation threshold of the conducting phase, yielding large ohmic
losses that increase the absorptance above
that of a corresponding homogeneous film.
Optimizing the thickness of the film to maximize its transmittance
above the percolation threshold of the conductive phase we
obtained a film with transmittance $T=0.41$ and a
sheet resistance $R_{\square}^{\mathrm{max}}\approx2.7\Omega$. We also
analyze the observed   
emission frequency shift of porous silicon electroluminescent devices
when Ag films are used as solid electrical contacts in replacement of
electrolytic ones.
\end{abstract}

\maketitle

\section{Introduction}

Transparent electric contacts are needed in a widespread variety of
optoelectronic applications. Materials such as
coinable metals are very good electrical conductors when compared to
the conductive polymers or semiconductor materials often used for those
applications,\cite{Ginley(2010)} but unfortunately, they are
opaque.
Organic conductive materials and doped metallic
oxides have been considered a good compromise, behaving as conductors
at low frequency and as dielectrics at optical
frequencies.\cite{Ginley(2010)} Nonetheless, the properties of
these semiconductor materials are defined by their chemical
composition and their doping. Thus, tuning the threshold
frequency at which they change behavior from conductor to dielectric
is difficult. An alternative for the 
design of semitransparent electrical contacts is the
use of nanostructured metallic-dielectric composites.
Extraordinary optical transmission\cite{Ebbesen(1998)} in 
perforated metallic films with nanoperforations has been explained in
terms of bulk and 
surface plasmons in nanostructured films\cite{Zayats(2003)} and is
promising for the control of optical
properties.\cite{Cai(2009)} These systems,  are composed of
a metallic and a dielectric phase; one of them may be
described as an array of nanometric inclusions with a given
geometry. Material composition, geometry, and 
order, affect the optical
properties of the system.\cite{Cortes(2010), Mochan(2010)} Fabrication of
nanostructured films with ordered patterns designed to tune optical
properties in the visible (VIS) range
requires high resolution lithography, employing 
interferometry of electronic
beams\cite{Brueck(2005),Martinez(2009)} or similar techniques. A
relative simple alternative is to use random
composite films.\cite{Brendan(2008)} Unlike the metallic oxides, this kind
of 
semitransparent contacts do not require high temperatures, are
flexible, might possess low enough surface roughness for the optical
range, and have a relatively low cost of production.

It is well known that the optical transmission\cite{Brendan(2008)} as
well as the electrical 
resistivity\cite{Maaroof(2005)}  of uniform metallic
thin films increase as the films becomes
thinner. However, the
reduction of the thickness of a film usually modifies its morphology
leading to inhomogeneities. A thin enough film is made of 
separate {\em islands}\cite{Katyayani(2003)} and is therefore
non-conducting. Near but above the
percolation threshold, while the conductive phase is
connected, there appear optical resonances at which the transmittance is
suppressed due
to the power dissipated as Joule heat.\cite{Toranzos(2010)} In the
present work, we employ a computationally efficient recursive
formalism for the calculation of an effective dielectric response of
nanostructured films when the length-scale of the inhomogeneities
of the film are much smaller than the
wavelength, thus neglecting
retardation.\cite{Cortes(2010),Mochan(2010),Mochan(2016)}  This
non-retarded recursive 
method (RM) is applicable \cite{Ortiz(2014)} to
nano-textured inhomogeneities 
with scales up to one order of magnitude below the nominal
wavelength. Analyzing
the optical and 
electrical properties of semicontinuous Ag films with a graded thickness
we have searched for an optimum film, with an adequate
conductivity in the low frequency range and a relatively high transmittance in
the VIS.

We also study Ag electrical
contacts on porous silicon (PS) electro-luminescent devices
(ELD).
It is known\cite{Canham(1997)} that in metal/PS/cSi junctions
under a bias voltage the injected carriers may
recombine. Due to the quantum confinement within the
thin Si regions in PS, direct 
radiative transitions with an energy larger than the bulk indirect gap
become permitted, leading to electro-luminescence (EL) in the VIS range of the
spectrum. 
Photo luminescence (PL) may also be observed if the sample is
irradiated by ultraviolet (UV) light. A relation between the PL spectra with
the porosity $p$ and the morphology of PS was
proposed by Bessa\"{\i} et al.\cite{Bessais(1996)}, yielding a maximum
emission at at 680nm with $p$=0.8. Under
similar preparation conditions, the PL and EL spectra are expected to
be similar\cite{Shi(1993)}. EL spectra from PS excited through 
an electrolytic contact\cite{Billat(1995)} have been characterized as
a function of the excitation potential and electric current, allowing
the control of the emission spectra through the preparation condition.
However, it has been found that the EL spectra of similarly prepared
PS films samples differ when they are excited through different solid
contacts. For PS with an expected porosity around 80\% over $p$ doped
Si with  Au vacuum evaporated contacts (Au/PS/$p$-Si/Al), a peak
emission was obtained at 680.\cite{Koshida(1992)} Similarly for an
ITO contact\cite{Koshida(1992)} or for an Al contact and an
$n$ doped substrate (Al/PS/$n$-Si/Al).\cite{Shi(1993)}. Nevertheless,
for an Au contact prepared by sputtering (Au/PS/$p$-Si/Al) the EL peak
shifts to 560nm\cite{Paredes(2002)}. Other shifts have been reported
for contacts made through the co-evaporation of Au and Ga or Sn
(530nm),  Au and In (455nm), and Au and Sb
(700nm).\cite{Steiner(1996)}
In the above cases the intensity of the EL signal was strongly
suppressed.

In this paper we also explore experimentally and theoretically the
spectral shift and the intensity suppression of the EL signal when
an electrolytic contact over an PS-ELD is replaced by a solid Ag
contact. 

The paper is organized as follows:
In section~\ref{sec:metodologia} we present our method
for fabricating Ag films with variable thickness
in only one evaporation step (subsection~\ref{subsec:fabrica}), and
we discuss our procedures for measuring the transmittance, ellipsometric
coefficients and resistance (~\ref{subsec:opticalandelectrical}). In
section~\ref{sec:Modelado} we 
discuss our model for the semicontinuous film and we describe our
computational procedures. In section~\ref{sec:Results} we obtain 
theoretical results from the RM for the ellipsometric
coefficients (\ref{subsec:ellipso}), the electrical
properties(\ref{subsec:conduct}),  and the
transmittance (\ref{subsec:transm})  of
semicontinuous films. As an application, in
section~\ref{sec:aplica} we 
fabricate
an optimally tuned solid transparent
electrical contact 
(\ref{subsec:espopt}) and we propose an explanation for the
spectral shift and the suppression of the observed emission of a PS-ELD
(subsec.~\ref{subsec:EL}). We devote section~\ref{sec:conclusiones} to
conclusions.

\section{Experimental}\label{sec:metodologia}

In this section we describe the experimental setup employed to
determine the optimal thickness of thin metallic films by fabricating
samples with a graded thickness and measuring their optical
transmittance, ellipsometric coefficients, and four-point electrical
resistance. The optimal thickness would be the one that maximizes the 
transmittance and minimizes the resistance.

\subsection{Sample Preparation}\label{subsec:fabrica}

\begin{figure}
\begin{center}
  \includegraphics[trim= 30 120 30 200,width=0.45\textwidth]{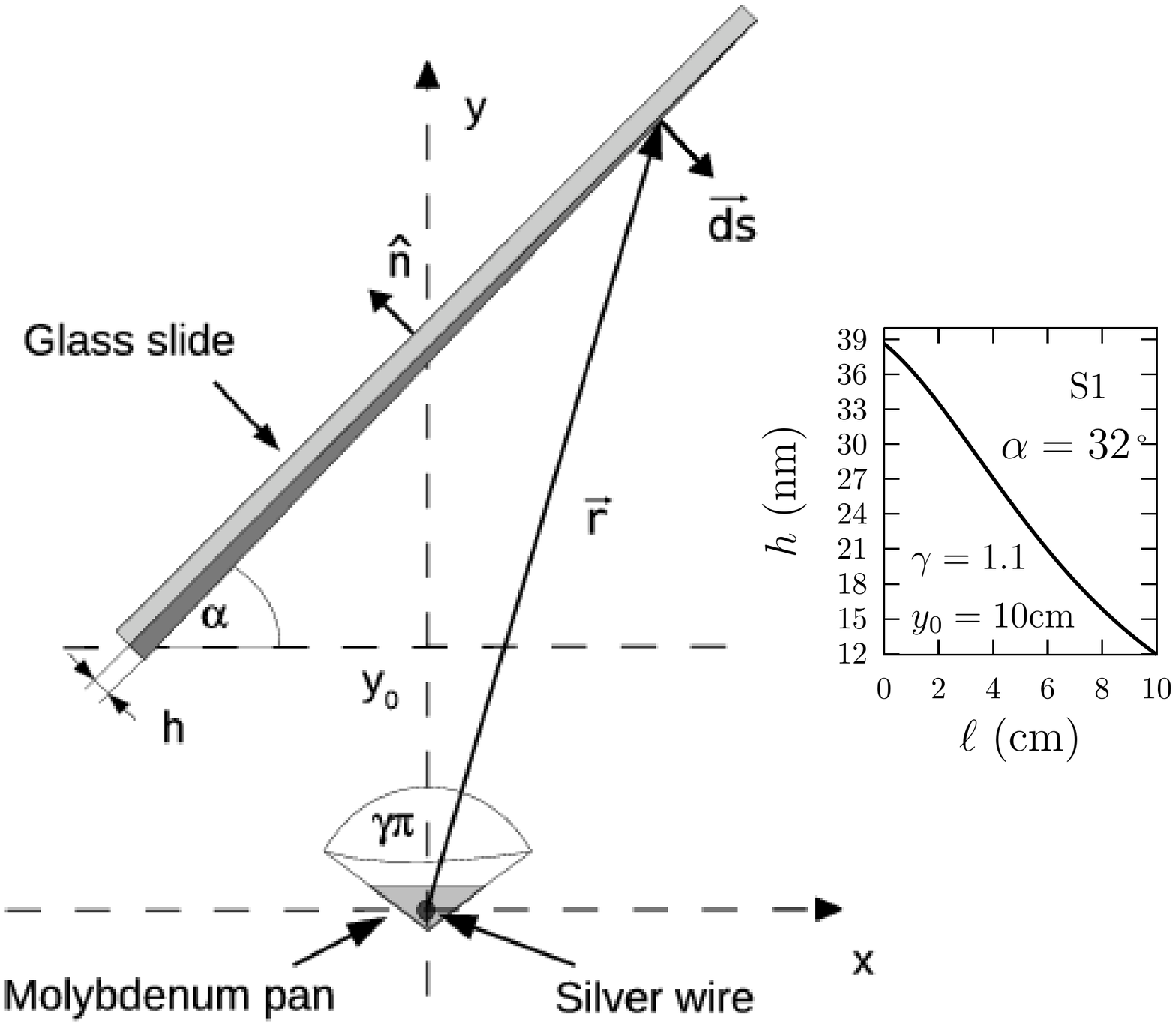}
\end{center}
\caption{ \label{fig:esq} Experimental setup for the growth of
  films with a graded thickness using a coinable-metal vacuum
  evaporation technique. A molybdenum crucible with an aperture that
  spans a solid 
  angle $\gamma\pi$ 
  supports a small piece 
  of Ag wire that is melted and vaporized by contact with an
  electrical heating device (not shown). The sample with 
  variable thickness (dark gray) is grown on a
  glass slide (light gray) tilted an angle $\alpha$. We indicate a differential
  element of surface $d\vec s=-ds\,\hat{n}$ at position $\vec r$, and
  the height $y_0$  of the lower edge of the slide. Inset: Nominal
  height $h$ versus 
  position  $\ell$ along the sample 
  calculated from $\alpha$, $y_0$, $m$ and $\gamma$ for sample $S1$.
}
\end{figure}

Fig.\ref{fig:esq} displays our setup for growing film samples with a
graded thickness ($h$), adapted from a
coinable-metal vacuum evaporation technique.\cite{Martin(2010), Bunshah(1994)}
The sample is obtained by isotropic thermal evaporation of Ag through
a solid angle $\gamma\pi$ determined by the aperture of a conical
molybdenum crucible whose walls screen the lower part of a vacuum bell
device. On a surface element $d\vec s$ at position $\vec r$ on a glass
slide, the deposited mass $dm$ is given by $-m\hat r\cdot d\vec
s/(\gamma\pi r^2)$ where $m$ is the total Ag mass vaporized.  A
nominal width $h$ may be obtained by dividing the mass per unit area
of the film by the bulk mass density of Ag. The actual density may be
smaller and the actual width larger due to texture of the film. Due to
the inclination 
$\alpha$ of the slide with respect to the horizontal, the deposited
film is thicker on the side that is closer to the source, at a height
$y_0$ above the crucible, and thinner
on the opposite side. Thus, we can prepare samples with
variable thickness within a range that depends on the geometrical
parameters of the experimental setup. We
show below results for two samples, both prepared with $\gamma=1.1$,
sample $S1$ with $y_0=$10 cm, $\alpha=32^\circ$, $m=16.5$mg and
sample $S2$ with  
$y_0=$9.7 cm, $\alpha=42^\circ$, and $m=8.25$mg. The inset of
Fig.\ref{fig:esq} shows the 
thickness $h$ as a function of position $\ell$ along the sample $S1$.

\subsection{Optical and electrical
  properties}\label{subsec:opticalandelectrical} 

We measured and averaged the optical normal-incidence transmittance
along three parallel equispaced  
lines running longitudinally through the sample in the
gradient direction, using (a)~a 650 nm
diode laser with a spot of diameter 1.5 mm, and (b)~an  Ocean Optics
UV-NIS-NIR source 
through an optical fiber. In both
cases we measured the intensity of the transmitted light at $\Delta
\ell=5$mm intervals. We used a clean 
glass slide as a reference. For (a) we used a detector
based on a photo-diode BPW20RF and in (b) the data were collected from
a USB spectrometer and we employed the {\em Spectrasuite} software. We also
determined the sheet resistance $R_\square$
over the same places using a four-point
technique.\cite{Martin(2010),Cohen(1983)}

We measured the ellipsometric parameters of the sample  $\psi$ and
$\delta$, defined through 
\begin{equation}\label{eq:fresnel}
  \frac{r_p}{r_s}=\exp(i\delta)\tan(\psi),
\end{equation}
where $r_p$ and $r_s$ are the reflection
coefficients for $p$ and $s$ polarization, respectively, 
with a Rudoph Research type 43702-200E ellipsometer in the
null field mode.\cite{Zerbino(2007)}  We fixed the incidence angle
$\theta$,
we put a linear polarizer and a quarter wave plate across the incident beam
oriented at angles $P$ and $C$ with respect to the plane of
incidence, respectively, and a linear analyzer across the
reflected beam at an angle $A$.
We set $C=\pi/4$ and found $\delta=2P+\pi/2$, and $\psi=A$ for different film
thicknesses and wavelengths, where we choose $P$ and $A$
to minimize the intensity of the outgoing 
beam.

\section{Model}\label{sec:Modelado}

For very thin films of coinable metals an {\em island}
morphology type has been reported.\cite{Maaroof(2005), Cai(2009)} As
the film grows the islands eventually connect among 
themselves, leading to a semicontinuous film  textured in the nanometric
scale.\cite{Bishop(2007),Maaroof(2005),Katyayani(2003)}  For
thickness smaller than a couple dozen nanometers, the air
interstices among the islands may sustain resonant excitations
that significantly modify the optical\cite{Cortes(2010),Mochan(2010)}
and electrical\cite{Ortiz(2011)} properties. As
we move along our graded sample towards its thick edge those
islands become more 
connected and for very thick films they merge into a continuous film.
\begin{figure}
  \includegraphics[trim= 20 140 50 140,width=0.45\textwidth]{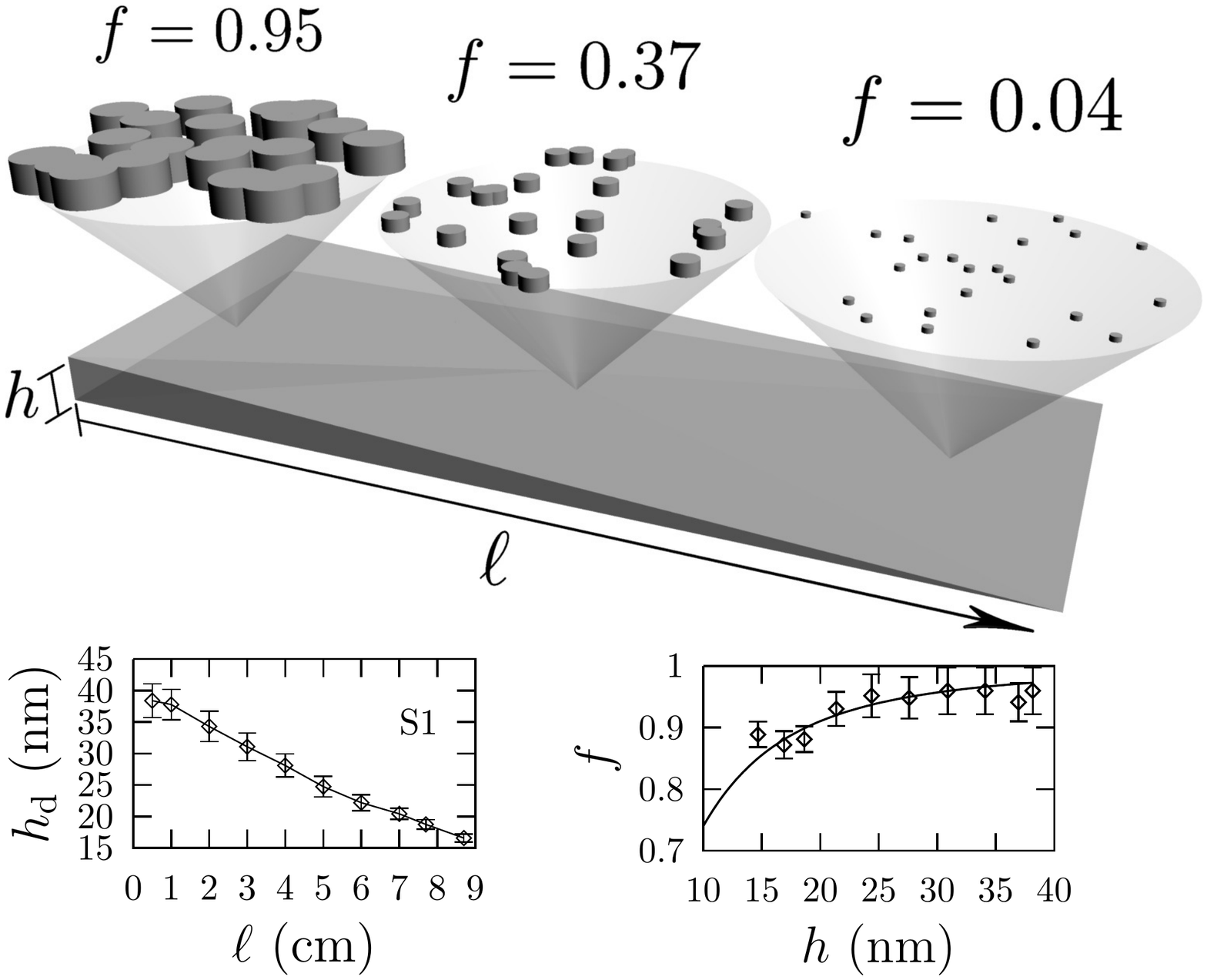}
  \caption{\label{fig:esquemaSample} Model film morphology at
    different positions $\ell$ along a sample prepared as in
    Fig. \ref{fig:esq}. The smooth wedge illustrates the nominal
    geometry, while the amplified regions illustrate our model,
    consisting of an ensemble of penetrable Ag
    disks of radius $a$ and height $h_d$ at uncorrelated random
    positions, yielding a semicontinuous film with filling fraction
    $f$. Below: Height of the disks $h_d$
    as a function of position $\ell$ and filling fraction $f$ as a
    function of nominal height $h$ as adjusted to the ellipsometric
    measurements (see text).
    }
\end{figure}
Fig. \ref{fig:esquemaSample} shows  
schematically the proposed morphology. Our model consists of an
ensemble of periodically repeated unit cells, each of which contains a
large enough number of penetrable disks of radius $a$
and height $h_d$
occupying random uncorrelated positions. The ensemble is characterized
by the filling fraction $f$, i.e., the fraction of the area covered by
the metal, and the amount of
metal deposited per unit area, which in turn is characterized by the
nominal height $h$. We employed the RM to 
calculate the macroscopic dielectric function $\epsilon^M$ of the
composite\cite{Cortes(2010),Mochan(2010),Ortiz(2014)} using the {\em
  Photonic} package.\cite{Mochan(2016)} The filling
fraction $f$ may be adjusted in the model by varying the 
radius $a$ of the disks and the height $h_d$ is related to $f$ through
$h_d=h/f$. 

\section{Results}\label{sec:Results}

\subsection{Ellipsometric coefficients}\label{subsec:ellipso}

\begin{figure}
    \includegraphics[trim= 20 300 20 200, width=0.45\textwidth]{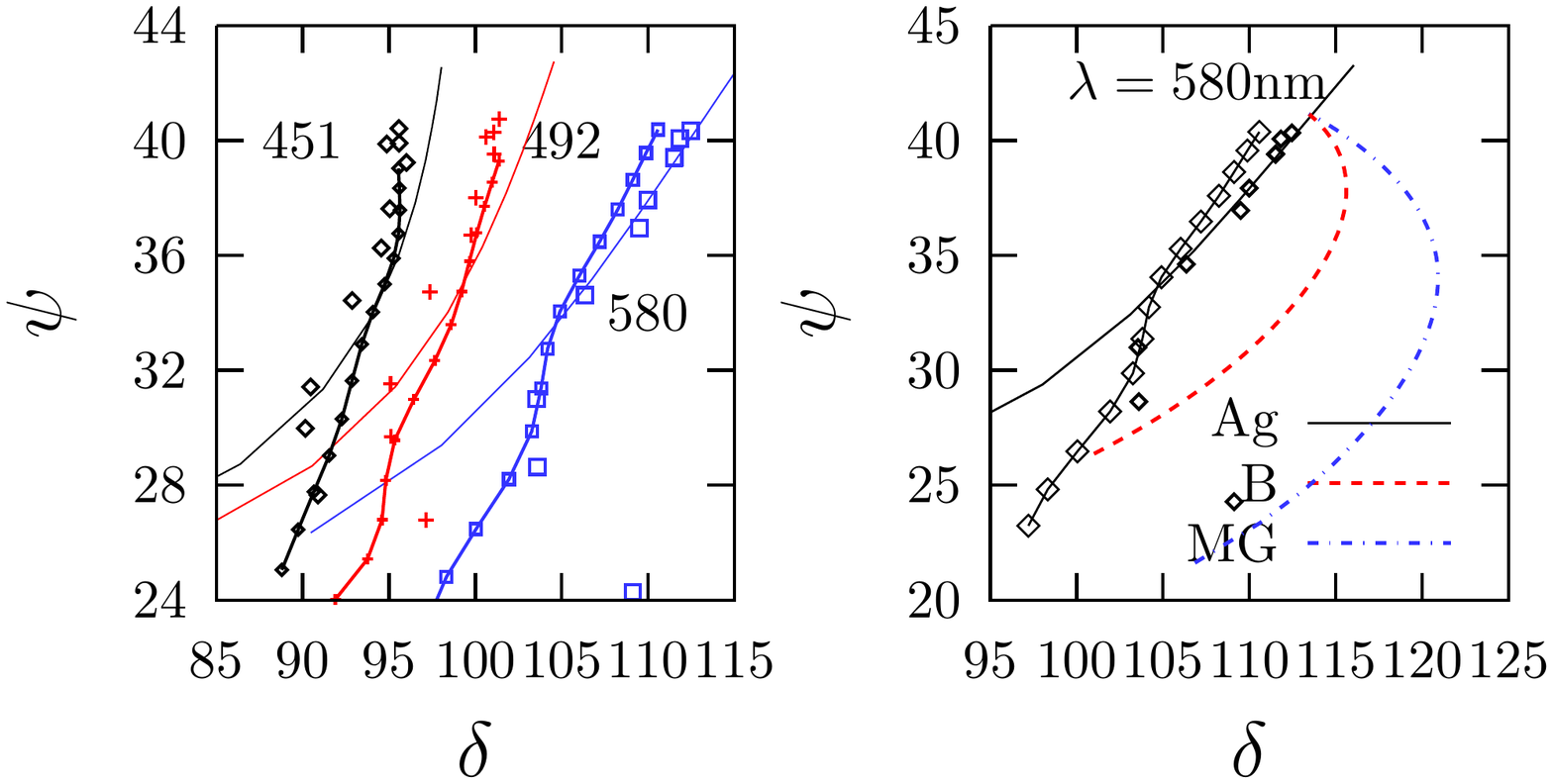}
    \caption{\label{fig:ellipso} Left panel: Ellipsometric coefficients
    ($\delta$, $\psi$) for three wavelengths $\lambda=$ 451
    ($\diamond$),
    492 ($+$) and 580 nm ($\square$) and for ten
    values of the 
    thickness $h$ 
    ($\psi$ increases with $h$ for each $\lambda$).
    The solid lines correspond to homogeneous Ag films, and the lines
    with points to the RM results. Right panel: Ellipsometric
    coefficients measured for a wavelength
    $\lambda=$580nm and ten different thicknesses ($\diamond$),
    calculated for homogeneous films (solid lines), with RM (points
    with lines), and with other effective media models: Bruggeman (B, 
    dashed lines) and Maxwell-Garnett (MG, dash-dotted lines).}
\end{figure}

We measured the ellipsometric coefficients at a fixed incidence angle
$\theta=69^{\circ}$.
We measured $\delta$ and $\psi$ at ten different positions $\ell$
along sample $S1$ 
and five wavelenghts $\lambda=$405, 451, 492, 546 and 580nm.
In the left panel of Fig. \ref{fig:ellipso} we show our results for
three of those (we omitted two to avoid cluttering the figure). 
Each point in Fig. \ref{fig:ellipso} corresponds to a different
thickness $h$ ($\psi$ increases with $h$). For thick films (above 20
nm)
experiment agrees roughly with the values of $\psi$ and $\delta$
calculated for a locally homogeneous film of width $h$ with a dielectric
function $\epsilon_{\mathrm{Ag}}$ taken from 
Ref. \onlinecite{Johnson(1972)} (commonly used for studying optical
properties of Ag composites at frequencies in UV-VIS-NIR range) and
deposited over glass, with dielectric response obtained from
Ref. \onlinecite{Palik(1985)}, but 
they don't agree for thinner films 
($h<20$nm) for which the film inhomogeneities lead to a strong
dependency of the optical response with the film morphology.
For each value of $\ell$ and $h$ we used our model (Subsec. \ref{sec:Modelado})
to obtain the macroscopic response of the film and identifying the
width of the film with the height $h_d$ of the disks we
calculated the 
ellipsometric coefficients. To this end we averaged our results over a
thousand realizations of our ensemble with thirty disks randomly
situated within a square unit cell. We have verified convergence of
the results. Our model has only one parameter, namely,
the radius $a$ of the disks, and it is adjusted for each value of
$\ell$ with corresponding nominal width $h$, to best reproduce the
ellipsometric measurements at the five wavelengths mentioned above. 
Fig.~\ref{fig:esquemaSample} displays
the fitted values of $h_d$ at the ten positions  $\ell$ for which we
measured $\delta$ and $\psi$. It also displays the resulting 
filling fraction $f$ as a function of the nominal height $h$,
together with an analytical fit, for which we chose the form
\begin{equation}\label{eq:fvsh}
  f=\frac{\mu h}{\sqrt{1+(\mu h)^2}},
\end{equation}
as it is linear for thin films and saturates
at $f=1$ for very thick ones. We obtained the parameter
$\mu=0.11\mathrm{nm}^{-1}$. 

The left panel of Fig. \ref{fig:ellipso} shows that results of our
RM calculation agree with experiment for most of  the film
thicknesses explored. In contrast, the effective medium models of
Maxwell-Garnett (MG) and Bruggeman (B)\cite{Etopim(1977)} differ
strongly, as shown in the right panel of
Fig. \ref{fig:ellipso} corresponding to $\lambda=$580nm.
Thus, our model system and the RM
computational procedure are much better suited for the calculation of
the effects of the nanometric texture on optical properties of very thin
films. It has
been reported\cite{Katyayani(2003),Maaroof(2005)}  that optical
properties of inhomogeneous films (such as 
semicontinuous film of coinable metals) differ from those of
homogeneous films. However, to our knowledge, this is the first time that
semicontinuous Ag films are analyzed and that the differences between
homogeneous and inhomogeneous films are calculated and compared to
those measured on a single sample consisting of a film with a graded 
thickness.  This kind of analysis allows
correlation of the parameters needed to design and manufacture semicontinuous
films with optimal parameters.

\subsection{Resistance}{\label{subsec:conduct}}

\begin{figure}
  \includegraphics[]{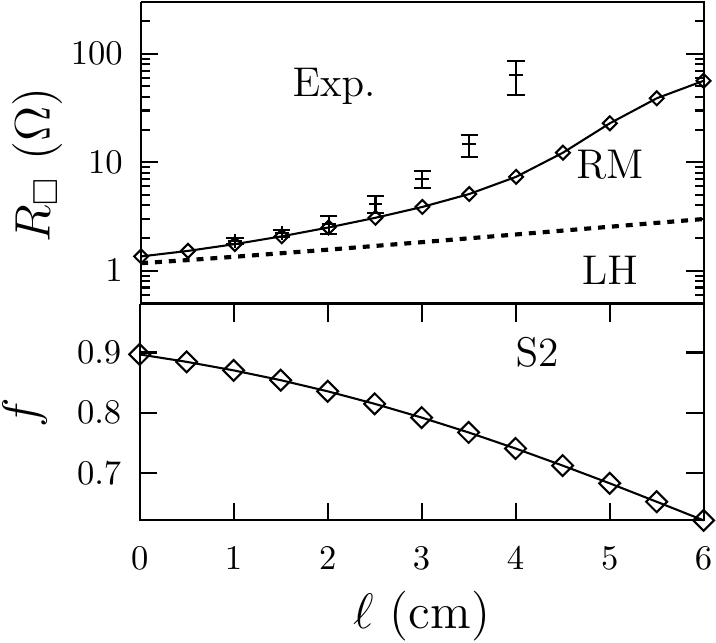}
  \caption{\label{fig:resistividad}
  Sheet resistance $R_{\square}$ measured for sample $S2$ (crosses
  with error bars) as a
  function of position $\ell$ along the sample (top panel). We also
  show the results 
  $\rho^M/h_d$ calculated with the RM theory for the morphology
  illustrated in Fig. \ref{fig:esquemaSample}
  (diamonds with lines) and the results $\rho_{\mathrm{Ag}}/h$ for a
  locally homogeneous Ag film (dashed line). In the bottom panel we include the
  calculated filling fraction $f$ as a function of position $\ell$.}
\end{figure}

In Fig. \ref{fig:resistividad} we show our measurements of the sheet
resistance $R_\square$ at various positions $\ell$ on sample
$S2$, prepared with a larger angle $\alpha$ and a smaller mass $m$
than sample $S1$, so that for each position $\ell$ the corresponding 
film is thinner. We also show the filling fraction
calculated for each $\ell$, using Eq. (\ref{eq:fvsh})
(fitted to the ellipsometric parameters of sample $S1$)
and using the
nominal height $h$ as a function of $\ell$ as described in 
Subsec. \ref{subsec:fabrica}. Notice that the resistance increases
very fast as $\ell$ increases, 
the film becomes thinner, and the filling fraction of the metal
diminishes, indicative of the approach to a percolation
transition at $f_c\approx0.7$ where the resistance
would diverge (we
couldn't measure the resistance for filling fractions smaller than
$0.74$). In
sample $S2$ this region corresponds roughly to the center of the
sample, whereas in sample $S1$ it is too close to one of its edges,
difficulting its measurement.

From the filling fraction $f$ and the nominal height $h$ we obtained
the height $h_d$ of our model disks. This allows us to
calculate the resistivity 
\begin{equation}
\label{eq:rho}
\rho^M=\frac{4\pi}{\mbox{lim}_{\omega \to
    0}\omega\mbox{Im}\left\{\epsilon^M(\omega)\right\}},
\end{equation}
and the sheet resistance $R_\square=\rho^M/h_d$ using the RM model.
To that end, we extrapolate the dielectric
function of the metallic 
phase towards low frequencies using the Drude model
\begin{equation}
\label{eq:dde}
\epsilon_{\mathrm{Ag}}= 1 - \frac{\omega^2_p}{\omega^2 + i\omega\gamma}.
\end{equation}
with parameters $\hbar\gamma=0.021$ eV,\cite{Cai(2009)}
$\hbar\omega_p=8.51$ eV, which correspond to the resistivity
$\rho_{\mbox{Ag}}=2.16\times10^{-6} \Omega$cm which we measured for
the same Ag wire from which the samples were prepared. 

The theoretical
results are displayed in Fig. \ref{fig:resistividad}. Notice that they
also increase rapidly as $\ell$ increases, though not as
rapidly as experiment. We recall that the geometrical percolation
threshold for penetrable disks is $f_c=0.676$\cite{Quintanilla(2000)}
and that our calculations are done on a system made up of a
periodically repeated finite random unit cell so that we expect finite
size effects to wash away the percolation transition in our
calculation. In the same 
figure we also include the sheet resistance $\rho_{Ag}/h$ of a
continuous film. Its 
behavior is qualitatively different from both experiment and our
calculation. Of course, for small $\ell$ and high filling fractions
both models coincide and agree with experiment.

\subsection{Transmittance}{\label{subsec:transm}}

\begin{figure}
\begin{center} 
  \includegraphics[width=0.45\textwidth]{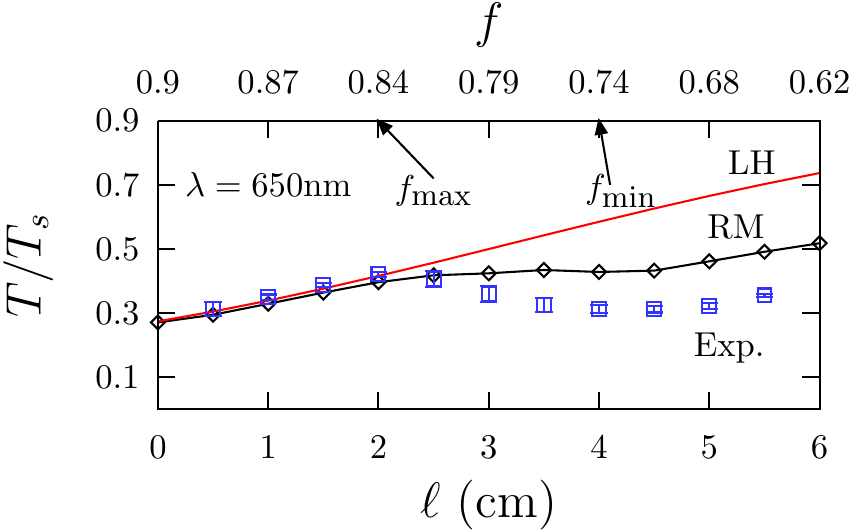}
\end{center}
\caption{\label{fig:tr650}  Transmittance $T$ versus position $\ell$
  along the sample $S2$ at wavelength $\lambda=650$nm normalized to
  the transmittance  $T_s$ of the 
  substrate.  We show experimental (Exp) results and results
  calculated through the RM theory (RM) for the morphology illustrated
  in Fig. \ref{fig:esquemaSample} and for a (locally) homogeneous film
  (LH). The corresponding filling fractions are indicated in the 
  top axis.}
\end{figure}

Fig.~\ref{fig:tr650} displays the transmittance $T$ normalized to the
transmittance $T_s$ of the glass slide as function
of position $\ell$, and thus, as a function of the filling fraction
$f$, measured at $\lambda$=650 nm for sample $S2$. We also 
show the transmittance calculated with the RM model and the result
for a homogeneous film. 

As $\ell$ increases and $f$ decreases the experimental transmittance
increases up to a local maximum $T=0.41$ corresponding to
a filling fraction $f_{\mathrm{max}}\approx0.84$ for which the
film is conducting as it lies above the
percolation threshold $f_c\approx0.7$ 
displayed in Fig. \ref{fig:resistividad}. Afterwards,  $T$
diminishes and reaches a local minimum
$T=0.29$ at $f_{\mathrm{min}}\approx0.74\gtrsim f_c$. As the filling 
fraction diminishes further, the transmittance of the film increases
again, but for $f<f_c$ the film is no longer
conducting. The RM calculation does not show the maximum and minimum
discussed above, but displays a similar inflection, and it
follows the experimental results much more closely than the
calculation for a homogeneous film.

The optical properties
close to the percolation threshold depend strongly on the morphology and
on the wavelength. In
this region field fluctuations are very intense and localized optical
resonances known as {\em hotspots} occur.
\begin{figure}
  \includegraphics[width=0.55\textwidth]{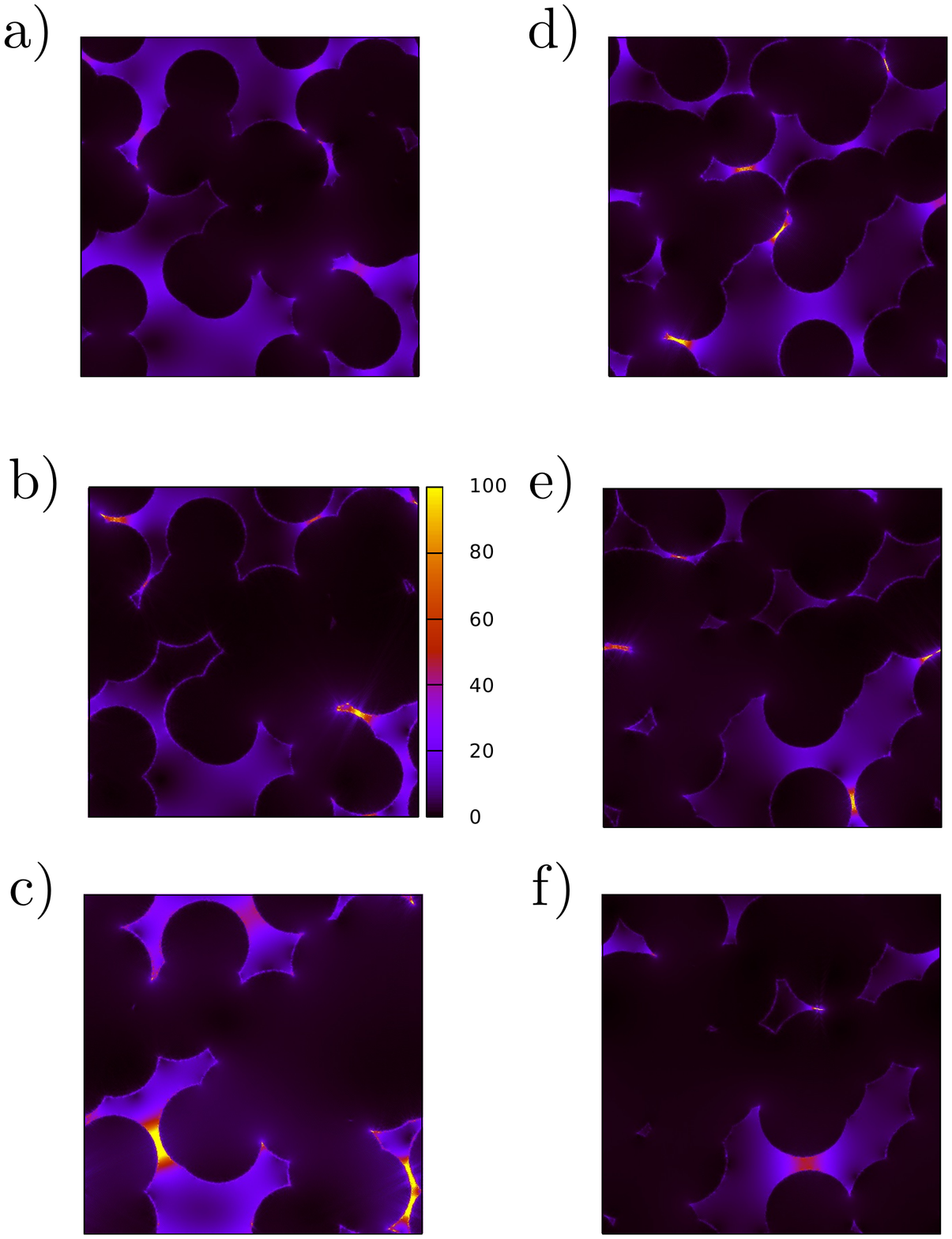}
\caption{\label{fig:hotspot} Absolute value $|\vec E|$ of
  the microscopic field normalized to that of the incident field
  $|\vec E_0|$ calculated with the RM model at
  $\lambda=650$nm for two members (left and right columns) of the
  ensemble and for three different radii and average filling fractions
  $a=0.11$, $f\approx0.69$ (a, d), $a=0.12$, $f\approx0.75$ (b, e),
  and $a=0.13$, $f\approx0.81$ (c, f).}
\end{figure}
To illustrate the hotspots, in Fig. \ref{fig:hotspot} we show the
microscopic field obtained through a recursive procedure based in the
RM calculation of the response.\cite{Mochan(2015),Mochan(2016)} We
show the field for two different members of the random ensemble that
models the morphology of the system,
and for three disk radii $a=0.11$, 0.12, 0.13, corresponding to the 
averaged filling fractions $f\approx0.69$, 0.75, 0.81, at a single wavelength
$\lambda=650$nm. 
Fig.\ref{fig:hotspot}c) displays hotspots in the interstice between two nearby 
particles (bottom left) and between three particles (bottom right), in
which $|\vec 
E|$ is approximately 2 orders of magnitude larger than its mean value.
Similar hotspots appear at different positions for different
wavelengths and for different filling fractions, as illustrated by the
other panels of the figure. Fig.\ref{fig:hotspot}b) shows four
different hotspots for a slightly smaller $f$. With $f$ even smaller 
Fig.\ref{fig:hotspot}a) shows no hotspots. But, another
member of the ensemble ensemble with the same $f$ displays several as
exemplified by Fig.\ref{fig:hotspot}d). The set of hotspots for that
member also change position and intensity as $f$ grows
(Figs. \ref{fig:hotspot}e and \ref{fig:hotspot}f). There are
well known results on the scaling of the localized states  
for disordered
random\cite{Shalaev(1996),Markel(1999)} and
fractal\cite{Stockman(1996),Ortiz(2001)} systems. These localized
states have found 
applications due to the corresponding enhancement in linear and
nonlinear signals. For example, SERS
and KERS enhancements of 1 to 2 dozen order of magnitude
have been reported.\cite{Moskovits(1985),Sarychev(2000)} 

\begin{figure}
\begin{center} 
  \includegraphics[width=0.45\textwidth]{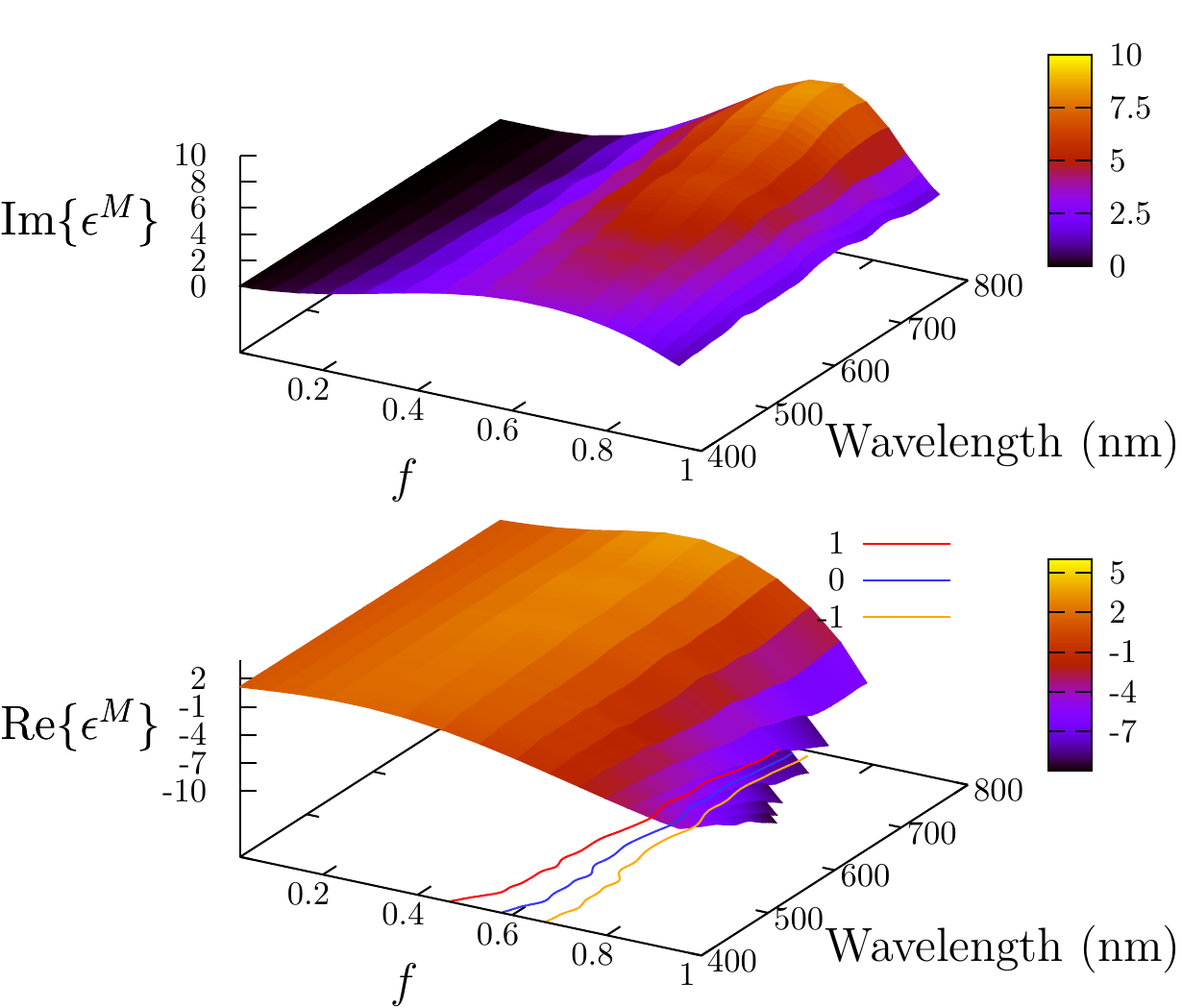}
\end{center}
\caption{\label{fig:epsMim} Ensemble average of the imaginary part
  (top panel) and real part (bottom panel) of 
  $\epsilon^M$ calculated within the RM 
  as a
  function of the filling fraction $f$ and the wavelength
  $\lambda$.}
\end{figure}

The presence of hotspots in the microscopic field for filling fractions
around the
percolation threshold is responsible for an increased energy
absorption within the conducting film, and thus to the decrease in the
transmittance $T$ as compared to that of a homogeneous film, as seen in
Fig.\ref{fig:tr650} for $f<f_{\mathrm{max}}$. This energy absorption corresponds
to Joule heat that is related to the imaginary part
of $\epsilon^M$.  Fig.\ref{fig:epsMim} displays the real and imaginary
parts of the ensemble average of $\epsilon^M$ as a function of the filling
fraction and of the wavelength.
For $f\lesssim1$, $h_d\approx h$, $\mathrm{Im}\{\epsilon^M$\} is
relatively small  
and $\mathrm{Re}\{\epsilon^M\}<0$.  Thus, the index of refraction
$n^M=\sqrt{\epsilon^M}$ is close to an  
imaginary number, the film is opaque and the field decays across it by a factor 
$\exp\{-\mbox{Im}(n^M) h\}$. Therefore, as $\ell$ increases and $h$
decreases the transmittance $T$ increases. As $f$
decreases further, $\mathrm{Re}\epsilon^M$ becomes less negative and
the field penetrates 
more into the film. Nevertheless,  $\mathrm{Im}\epsilon^M$
also increases and reaches a maximum for $f\approx0.72$, thus
increasing the absorption and decreasing the transmittance. For even
larger $\ell$ and smaller $f$ the film becomes dielectric and the
transmittance increases again. This explains qualitatively the maximum
and minimum transmittance observed in Fig. \ref{fig:tr650}. The
corresponding inflection in the transmittance is reproduced by our
numerical calculation, although not the actual maximum and minimum.

\section{Applications}\label{sec:aplica}

\subsection{Optimization of semicontinuous Ag film}\label{subsec:espopt}

\begin{figure}
\begin{center} 
  \includegraphics[width=0.45\textwidth]{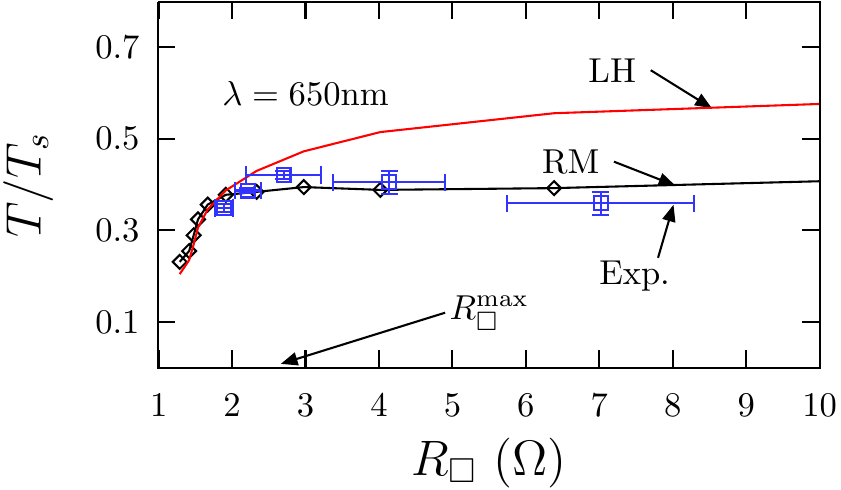}
\end{center}
\caption{\label{fig:r3} Transmittance $T$ of a film normalized to the
  transmittance $T_s$ of the glass substrate vs. the sheet resistance 
  $R_\square$ for $\lambda$=650 nm. We show experimental results as
  well as the results from the RM calculation and results for a
  locally homogeneous film.
  }
\end{figure}

An optimal transparent electrical contact would have the highest
possible transmittance and the lowest possible resistance. To choose
the best combination of parameters we plot  $T$ as a function of
$R_\square$ in Fig. \ref{fig:r3}, combining information from
Figs. \ref{fig:resistividad} and 
\ref{fig:tr650}.  
It is clear that it is not useful to diminish the nominal width $h$ of the film
too much, as the transmittance stops increasing while the
resistance does. Thus, an optimal choice would correspond to the
maximum transmittance which corresponds to $T/T_s=0.41$, $R=2.7 \Omega$,
$\ell=2$ cm, $f=0.84$, $h=13.8$ nm, $h_d=16.4$ nm 

\subsection{Electro-luminescence of PS with an Ag contact}\label{subsec:EL}

We analyze light emitting devices (LED) made of porous silicon (PS)
over which Ag is sputtered to form an electrical contact. We prepare
the PS sample by 
anodizing a crystalline $p$-doped Si substrate with resistivity
0.5-1.0$\Omega$cm terminated in a (100) surface, using a well known
procedure:\cite{Pavesi(2003)} We immerse the Si crystal in an electrolytic
solution of hydrofluoric acid, distilled water, and ethanol
in proportions 1:1:2 and we apply a current with a density of 20
mA/cm$^2$ for a duration of 120 s\cite{Toranzos(2010)} to obtain a
sample with an expected porosity $p\approx80\%$. We 
discard the electrolytic solution and clean the system with an ethanol
bath.

\begin{figure}
\begin{center} 
  \includegraphics[width=0.4\textwidth]{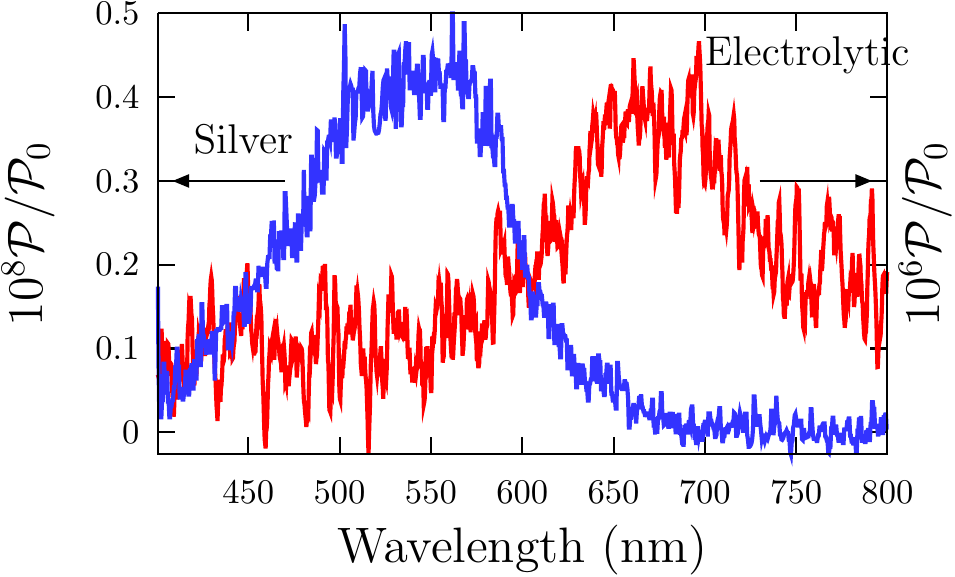}
\end{center}
\caption{\label{fig:LED} Power  $\mathcal P$ emitted by a porous Si
  light emitting device as a function of wavelength $\lambda$,
  normalized to the electrical input power $\mathcal P_0$. The peak on the
  right (right scale) corresponds to an electrolytic contact of
  diluted KCl and a current density of 16mA/cm$^{2}$. The peak on the
  left (left scale) corresponds to a 
  contact made of a semicontinuous silver film and a current density
  of 20mA/cm$^{2}$.}
\end{figure}

If we use a diluted solution of potassium chloride as an electrolytic
contact and apply a forward current with density 16 mA/cm$^2$ an
electroluminescent signal is produced. We measured the normalized
spectral emission with an Hitachi fluorimeter F2000 with blocked
excitation source $25$s after turning the current
on\cite{Toranzos(2008)b}. The results are displayed in
Fig.~\ref{fig:LED}

It is well known that luminescence of PS\cite{Buda(1992)} is due to the
confinement of charge carriers within the pore walls, which introduces
an uncertainty in the momentum that allows direct transitions
that would be forbidden in bulk Si.\cite{Kovalev(1999)} Thus, the
$\lambda=1130$ 
indirect gap of bulk Si gives rise to a direct electronic transition in PS
that is shifted due to quantum confinement and corresponds to 
the emitted emission with wavelength $\lambda=680$nm illustrated by the peak on
the right of Fig.\ref{fig:LED} corresponding to an electrolytic
contact.\cite{Steiner(1996)} While the current flows, silicon
oxide is progressively produced on the walls of the pores, increasing
the confinement and producing a small blue shift of the maximum emission
peak.\cite{Pavesi(2003)}
After the current had been applied for approximately $90$s the
electroluminiscence of our sample was extinguished.

The electrical excitation of PS may also be realized with 
Schottky type metallic solid
contacts.\cite{Steiner(1996)} An advantage of metallic contacts is that
the electric current doesn't produce the chemical reactions that
rapidly destroys the electroluminiscence when an electrolytic contact
is employed, extending the life of the device. 
To produce a metallic contact,
we use cathodic projection (sputtering) of Ag
on a $0.2\mathrm{cm}^2$
masked surface of a PS sample fabricated as described above.
We chose the sputtering conditions (pressure,
arc current, distance of Ag target to sample, deposition time)
by first sputtering onto a glass sample and selecting those conditions
that yielded a film with relative
transmittance $T/T_s\approx0.2$.

Fig.~\ref{fig:LED} 
shows the electroluminiscence spectrum of our sample with a metallic
contact excited by an electric current with density
density of 20 mA/cm$^2$. Comparing this spectrum with that
corresponding to the electrolytic contact, we notice that it is
strongly blue-shifted towards the central part of VIS spectra
($\approx$ 550nm),  but that its intensity is about  
two orders of magnitude smaller.

\begin{figure}
\begin{center} 
  \includegraphics[width=0.45\textwidth]{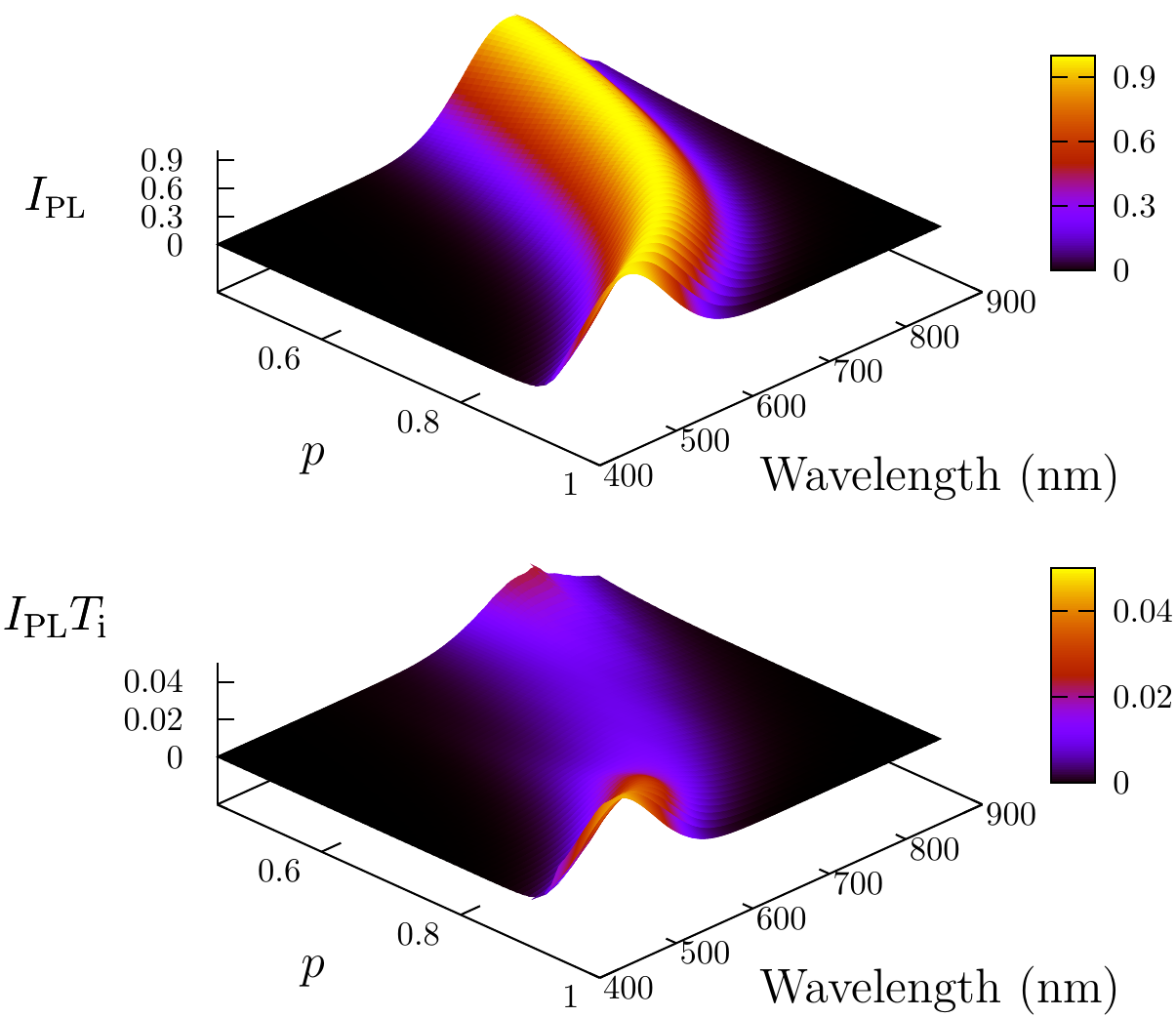}
\end{center}
\caption{\label{fig:trLed} Top panel: Normalized photoluminiscence signal
  $I_{\mathrm{PL}}$ of porous silicon (PS) calculated as a function of porosity
  $p$ and wavelength $\lambda$.\cite{Pavesi(2003)} Bottom panel:
  Product $I_{\mathrm{PL}}$ of top panel with
  $T_{\mathrm{i}}$ for a simulated composite 
  PS and PS permeated with Ag.}
\end{figure}

To understand the shift of the maximum of the emission spectra from
$\lambda=680$ for an 
electrolytic contact to $\lambda=550$ for an Ag contact, we obtained
the emission spectra from the photo-luminescence (PL)
spectra of Si nano-crystals and their size
dependence.\cite{Kovalev(1999)} The effects of confinement in the
walls of PS have been related to the effects of confinement in
nano-crystals and a relation between porosity and effective size has been
proposed,\cite{Bessais(1996)} as well as analytical expressions for
the shift of the emission peak with respect to
size.\cite{Fauchet(1997)} If the sample is kept in air there is a 
further blue shift due to oxidation at the surface of the
pores.\cite{Paredes(2002),Steiner(1996)} Thus, we model the intrinsic
PL spectrum of PS as a Gaussian of an appropriate width centered at
the energy corresponding to the porosity, and we assume that the
EL emission spectrum is similar to the PL one.
In the top panel of Fig.\ref{fig:trLed} we show the resulting
PL spectra as a function of both wavelength and
porosity.\cite{Pavesi(2003)}.

When using a solid contact, the light produced through EL is
transmitted through the contact before being observed. Thus we expect
that the observed EL signal should be given by the product of the
emission, similar to the the PL spectrum, and the optical transmittance of the
system PS/contact/air. We model the contact by assuming most of the
metal infiltrates the pores. 
We used the RM method to
calculate the macroscopic dielectric response of PS modeled as a
random ensemble of empty cylinders within a crystalline Si host of
porosity $p$,\cite{Ortiz(2011)} and we also use the RM method to
calculate the macroscopic dielectric response of the contact modeled
as a random ensemble of Ag cylinders within the crystalline Si host, with
a filling fraction given by the porosity. 
We calculate the transmittance 
$T_{\mathrm{i}}$ of the system formed by PS covered
with a film of width $d$ made up of Ag-infiltrated PS.
In Fig.\ref{fig:trLed} we also show the product of the intrinsic
PS luminescence signal with the transmittance
$T_{\mathrm{i}}$ for $d=45$nm. Notice that the signal is suppressed
about two orders of magnitude with respect to the PL spectrum and that
it is becomes blue shifted and more intense as the porosity
increases.

\begin{figure}
\begin{center} 
  \includegraphics[width=0.45\textwidth]{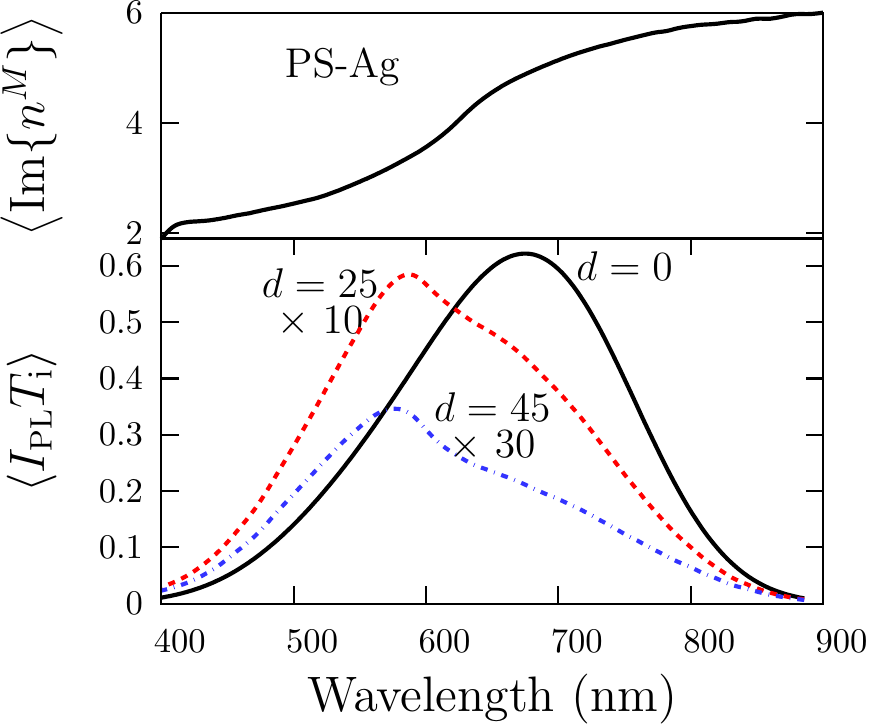}
\end{center}
\caption{\label{fig:LEDp08} Bottom panel: EL spectra calculated by
  multiplying the luminescence $I_{\mathrm{PL}}$ of PS and the
  transmittance $T_{\mathrm{i}}$ of the system PS/contact/air as function of
  wavelength $\lambda$ for different values of the thickness of the
  contact $d=0$nm, 25nm, and 45nm. The results are averaged over porosity $p$
  assuming a Gaussian distribution of width  $\sigma=0.14$ centered at
  $\bar p=0.8$. The curves for $d=25$nm and
  $d=45$nm are multiplied by 10 and 30 respectively. Top panel: 
  imaginary part of the index of refraction $n^M$ of the contact. }
\end{figure}

Finally, we assume that the porosity in our random sample is not fixed
at the nominal value $\bar p=80$\% expected from our preparation
procedure, but has 
some fluctuations. We assume $p$ is distributed
as a Gaussian of width $\sigma=0.14$ around its average $\bar p$ and. In
Fig. \ref{fig:LEDp08} we show the corresponding EL spectra for
different thicknesses $d$ of the contact. 

The curve for $d=0$ corresponds to the absence of metal,
and thus may be compared to the EL spectrum with a transparent
electrolytic contact. It has a peak at 680 nm which agrees well with
that in the right side of Fig. \ref{fig:LED}. The peak is slightly red
shifted with respect to the nominal peak corresponding to $p=\bar
p=0.8$ in the top panel of Fig. \ref{fig:trLed}, as the direction of
its ridge is closer to that of the $p$ axis for lower frequencies. 

On the other hand, the peaks of the curves corresponding to 
$d=25$nm and $d=45$nm are blue shifted to 590nm and 570nm and are
suppressed about one and two orders of magnitude respectively. The
reason for the shift and the height decrease of the peaks lies in the
absorption within the contact, as illustrated in the top panel of
Fig. \ref{fig:LEDp08}, which shows that the macroscopic index of
refraction of the contact is large and increases with wavelength, therefore
yielding a stronger suppression at the red end of the spectrum.  Thus,
our model of a metallic contact that partially infiltrates the pores
PS is able to explain blue shift
and suppression of the EL spectra observed experimentally and
illustrated by the curve on the left side of 
Fig. \ref{fig:LED}. On the other hand, we were unable to reproduce the
experimental 
results using a metallic semicontinuous overlayer, as in the previous
sections, as a contact. Our results are not 
very sensitive to the width $\sigma$ of the porosity distribution, but
we don't get a large enough shift in the limit $\sigma\to0$.

\section{Conclusions}\label{sec:conclusiones}

We prepared Ag thin films with a height gradient by evaporation onto a
tilted glass substrate. We performed ellipsometric measurements which showed
than the films changed nature from locally homogeneous to
semicontinuous and to an island morphology as their width
diminished. We modeled the system as an ensemble of
penetrable metallic cylinders occupying random uncorrelated positions,
and we applied an efficient 
recursive formalism to obtain its macroscopic dielectric function. We
fitted the ellipsometric measurements to relate the position along the
sample film and the nominal height of the film to its actual height
and its (area) filling fraction. We measured and analyzed with the
previous fit the sheet resistance of the film and identified a
percolation transition where the film resistance along the film
diverges. The optical transmittance of the film was also measured and
analyzed, and we obtained a deviation from the transmittance of a
homogeneous film. As the film becomes narrower the transmittance
increases, reaches a maximum and then it decreases until it
hits a minimum and increases again, but this time as a dielectric and
not a conducting film. This behavior has been attributed to the
scattering of light by the inhomogeneities of the
film\cite{Bishop(2007)}. Nevertheless, these inhomogeneities are not
expected to have a lengthscale much larger than the height of the
film, much smaller than the wavelength of light, so that scattering is
expected to be small\cite{Bohren(1983)} and the film may be
characterized by a macroscopic effective permittivity.\cite{Ortiz(2014)}  
Our model calculation of the dielectric response captured qualitatively
the behavior of the transmittance, although with smaller oscillations
as the filling fraction decreases, displaying an inflection, but
without an actual maximum and a minimum. We explained 
qualitatively the behavior of the transmittance in terms of plasmonic
resonances within the film. For a range of filling fractions around
the percolation threshold a series of resonances appear for which the
field becomes very intense in very small regions. These hotspots are
responsible for an increased energy dissipation within the film, as
shown by an the imaginary part of the 
macroscopic dielectric function of the system. We employed our results
to optimize a film for use as a semitransparent contact. We also
calculated the shift towards the blue and the decrease in the
intensity of the electro-luminescence of porous silicon when a metallic
semicontinuous contact is employed instead of an electrolytic one, and
obtained agreement with experiment assuming that the metal infiltrates
the pores and that the porosity has some fluctuations around its mean.

\section*{Acknowledgment} 
We acknowledge support from ANPCyT/FONCyT and UNNE through the grants
PICT/2013-0696, PI-F008/2014-SGCyT, and from DGAPA-UNAM through grant
No. IN113016.
\bibliography{referencias} 

\begin{thebibliography}{44}%
\makeatletter
\providecommand \@ifxundefined [1]{%
 \@ifx{#1\undefined}
}%
\providecommand \@ifnum [1]{%
 \ifnum #1\expandafter \@firstoftwo
 \else \expandafter \@secondoftwo
 \fi
}%
\providecommand \@ifx [1]{%
 \ifx #1\expandafter \@firstoftwo
 \else \expandafter \@secondoftwo
 \fi
}%
\providecommand \natexlab [1]{#1}%
\providecommand \enquote  [1]{``#1''}%
\providecommand \bibnamefont  [1]{#1}%
\providecommand \bibfnamefont [1]{#1}%
\providecommand \citenamefont [1]{#1}%
\providecommand \href@noop [0]{\@secondoftwo}%
\providecommand \href [0]{\begingroup \@sanitize@url \@href}%
\providecommand \@href[1]{\@@startlink{#1}\@@href}%
\providecommand \@@href[1]{\endgroup#1\@@endlink}%
\providecommand \@sanitize@url [0]{\catcode `\\12\catcode `\$12\catcode
  `\&12\catcode `\#12\catcode `\^12\catcode `\_12\catcode `\%12\relax}%
\providecommand \@@startlink[1]{}%
\providecommand \@@endlink[0]{}%
\providecommand \url  [0]{\begingroup\@sanitize@url \@url }%
\providecommand \@url [1]{\endgroup\@href {#1}{\urlprefix }}%
\providecommand \urlprefix  [0]{URL }%
\providecommand \Eprint [0]{\href }%
\providecommand \doibase [0]{http://dx.doi.org/}%
\providecommand \selectlanguage [0]{\@gobble}%
\providecommand \bibinfo  [0]{\@secondoftwo}%
\providecommand \bibfield  [0]{\@secondoftwo}%
\providecommand \translation [1]{[#1]}%
\providecommand \BibitemOpen [0]{}%
\providecommand \bibitemStop [0]{}%
\providecommand \bibitemNoStop [0]{.\EOS\space}%
\providecommand \EOS [0]{\spacefactor3000\relax}%
\providecommand \BibitemShut  [1]{\csname bibitem#1\endcsname}%
\let\auto@bib@innerbib\@empty
\bibitem [{\citenamefont {Ginley}(2010)}]{Ginley(2010)}%
  \BibitemOpen
  \bibfield  {author} {\bibinfo {author} {\bibfnamefont {D.~S.}\ \bibnamefont
  {Ginley}},\ }\href@noop {} {\emph {\bibinfo {title} {Handbook of Transparent
  Conductors}}}\ (\bibinfo  {publisher} {Springer, USA},\ \bibinfo {year}
  {2010})\BibitemShut {NoStop}%
\bibitem [{\citenamefont {Ghaemi}\ \emph {et~al.}(1998)\citenamefont {Ghaemi},
  \citenamefont {Thio}, \citenamefont {Grupp}, \citenamefont {Ebbesen},\ and\
  \citenamefont {Lezec}}]{Ebbesen(1998)}%
  \BibitemOpen
  \bibfield  {author} {\bibinfo {author} {\bibfnamefont {H.}~\bibnamefont
  {Ghaemi}}, \bibinfo {author} {\bibfnamefont {T.}~\bibnamefont {Thio}},
  \bibinfo {author} {\bibfnamefont {D.}~\bibnamefont {Grupp}}, \bibinfo
  {author} {\bibfnamefont {T.}~\bibnamefont {Ebbesen}}, \ and\ \bibinfo
  {author} {\bibfnamefont {H.}~\bibnamefont {Lezec}},\ }\href@noop {}
  {\bibfield  {journal} {\bibinfo  {journal} {Phys. Rev. B}\ }\textbf {\bibinfo
  {volume} {58}},\ \bibinfo {pages} {6779} (\bibinfo {year}
  {1998})}\BibitemShut {NoStop}%
\bibitem [{\citenamefont {Darmanyan}\ and\ \citenamefont
  {Zayats}(2003)}]{Zayats(2003)}%
  \BibitemOpen
  \bibfield  {author} {\bibinfo {author} {\bibfnamefont {S.}~\bibnamefont
  {Darmanyan}}\ and\ \bibinfo {author} {\bibfnamefont {A.}~\bibnamefont
  {Zayats}},\ }\href@noop {} {\bibfield  {journal} {\bibinfo  {journal} {Phys.
  Rev. B}\ }\textbf {\bibinfo {volume} {67}},\ \bibinfo {pages} {035424}
  (\bibinfo {year} {2003})}\BibitemShut {NoStop}%
\bibitem [{\citenamefont {Cai}\ and\ \citenamefont
  {Shalaev}(2009)}]{Cai(2009)}%
  \BibitemOpen
  \bibfield  {author} {\bibinfo {author} {\bibfnamefont {W.}~\bibnamefont
  {Cai}}\ and\ \bibinfo {author} {\bibfnamefont {V.}~\bibnamefont {Shalaev}},\
  }\href@noop {} {\emph {\bibinfo {title} {Optical Metamaterials: Fundamentals
  and Applications}}}\ (\bibinfo  {publisher} {Springer, New York},\ \bibinfo
  {year} {2009})\BibitemShut {NoStop}%
\bibitem [{\citenamefont {Cortes}\ \emph {et~al.}(2010)\citenamefont {Cortes},
  \citenamefont {Moch\'an}, \citenamefont {Mendoza},\ and\ \citenamefont
  {Ortiz}}]{Cortes(2010)}%
  \BibitemOpen
  \bibfield  {author} {\bibinfo {author} {\bibfnamefont {E.}~\bibnamefont
  {Cortes}}, \bibinfo {author} {\bibfnamefont {W.~L.}\ \bibnamefont
  {Moch\'an}}, \bibinfo {author} {\bibfnamefont {B.~S.}\ \bibnamefont
  {Mendoza}}, \ and\ \bibinfo {author} {\bibfnamefont {G.~P.}\ \bibnamefont
  {Ortiz}},\ }\href@noop {} {\bibfield  {journal} {\bibinfo  {journal} {Phys.
  Stat. Sol. B}\ }\textbf {\bibinfo {volume} {247}},\ \bibinfo {pages} {2102}
  (\bibinfo {year} {2010})}\BibitemShut {NoStop}%
\bibitem [{\citenamefont {Moch\'an}\ \emph {et~al.}(2010)\citenamefont
  {Moch\'an}, \citenamefont {Ortiz},\ and\ \citenamefont
  {Mendoza}}]{Mochan(2010)}%
  \BibitemOpen
  \bibfield  {author} {\bibinfo {author} {\bibfnamefont {W.~L.}\ \bibnamefont
  {Moch\'an}}, \bibinfo {author} {\bibfnamefont {G.~P.}\ \bibnamefont {Ortiz}},
  \ and\ \bibinfo {author} {\bibfnamefont {B.~S.}\ \bibnamefont {Mendoza}},\
  }\href@noop {} {\bibfield  {journal} {\bibinfo  {journal} {Opt. Express}\
  }\textbf {\bibinfo {volume} {18}},\ \bibinfo {pages} {22119} (\bibinfo {year}
  {2010})}\BibitemShut {NoStop}%
\bibitem [{\citenamefont {Brueck}(2005)}]{Brueck(2005)}%
  \BibitemOpen
  \bibfield  {author} {\bibinfo {author} {\bibfnamefont {S.}~\bibnamefont
  {Brueck}},\ }\href@noop {} {\bibfield  {journal} {\bibinfo  {journal} {Proc.
  IEEE}\ }\textbf {\bibinfo {volume} {93}},\ \bibinfo {pages} {1704} (\bibinfo
  {year} {2005})}\BibitemShut {NoStop}%
\bibitem [{\citenamefont {Mart\'inez}\ \emph {et~al.}(2009)\citenamefont
  {Mart\'inez}, \citenamefont {Bellino},\ and\ \citenamefont
  {Soler-Illia}}]{Martinez(2009)}%
  \BibitemOpen
  \bibfield  {author} {\bibinfo {author} {\bibfnamefont {E.}~\bibnamefont
  {Mart\'inez}}, \bibinfo {author} {\bibfnamefont {M.}~\bibnamefont {Bellino}},
  \ and\ \bibinfo {author} {\bibfnamefont {G.}~\bibnamefont {Soler-Illia}},\
  }\href@noop {} {\bibfield  {journal} {\bibinfo  {journal} {ACS applied
  materials \& interfaces}\ }\textbf {\bibinfo {volume} {1}},\ \bibinfo {pages}
  {746} (\bibinfo {year} {2009})}\BibitemShut {NoStop}%
\bibitem [{\citenamefont {Connor}\ \emph {et~al.}(2008)\citenamefont {Connor},
  \citenamefont {Haughn}, \citenamefont {An}, \citenamefont {Pipe},\ and\
  \citenamefont {Shtein}}]{Brendan(2008)}%
  \BibitemOpen
  \bibfield  {author} {\bibinfo {author} {\bibfnamefont {B.~O.}\ \bibnamefont
  {Connor}}, \bibinfo {author} {\bibfnamefont {C.}~\bibnamefont {Haughn}},
  \bibinfo {author} {\bibfnamefont {K.}~\bibnamefont {An}}, \bibinfo {author}
  {\bibfnamefont {K.}~\bibnamefont {Pipe}}, \ and\ \bibinfo {author}
  {\bibfnamefont {M.}~\bibnamefont {Shtein}},\ }\href@noop {} {\bibfield
  {journal} {\bibinfo  {journal} {Appl. Phys. Lett}\ }\textbf {\bibinfo
  {volume} {93}},\ \bibinfo {pages} {223304} (\bibinfo {year}
  {2008})}\BibitemShut {NoStop}%
\bibitem [{\citenamefont {Maaroof}\ and\ \citenamefont
  {Smith}(2005)}]{Maaroof(2005)}%
  \BibitemOpen
  \bibfield  {author} {\bibinfo {author} {\bibfnamefont {A.~I.}\ \bibnamefont
  {Maaroof}}\ and\ \bibinfo {author} {\bibfnamefont {G.~B.}\ \bibnamefont
  {Smith}},\ }\href@noop {} {\bibfield  {journal} {\bibinfo  {journal} {Thin
  Solid Films}\ }\textbf {\bibinfo {volume} {485}},\ \bibinfo {pages} {198}
  (\bibinfo {year} {2005})}\BibitemShut {NoStop}%
\bibitem [{\citenamefont {Seal}\ \emph {et~al.}(2003)\citenamefont {Seal},
  \citenamefont {Nelson},\ and\ \citenamefont {Ying}}]{Katyayani(2003)}%
  \BibitemOpen
  \bibfield  {author} {\bibinfo {author} {\bibfnamefont {K.}~\bibnamefont
  {Seal}}, \bibinfo {author} {\bibfnamefont {M.~A.}\ \bibnamefont {Nelson}}, \
  and\ \bibinfo {author} {\bibfnamefont {Z.~C.}\ \bibnamefont {Ying}},\
  }\href@noop {} {\bibfield  {journal} {\bibinfo  {journal} {Phys. Rev. B}\
  }\textbf {\bibinfo {volume} {67}},\ \bibinfo {pages} {035318} (\bibinfo
  {year} {2003})}\BibitemShut {NoStop}%
\bibitem [{\citenamefont {Toranzos}\ \emph {et~al.}(2010)\citenamefont
  {Toranzos}, \citenamefont {Ortiz},\ and\ \citenamefont
  {Koropecki}}]{Toranzos(2010)}%
  \BibitemOpen
  \bibfield  {author} {\bibinfo {author} {\bibfnamefont {V.}~\bibnamefont
  {Toranzos}}, \bibinfo {author} {\bibfnamefont {G.}~\bibnamefont {Ortiz}}, \
  and\ \bibinfo {author} {\bibfnamefont {R.}~\bibnamefont {Koropecki}},\ }in\
  \href@noop {} {\emph {\bibinfo {booktitle} {ANALES AFA}}},\ Vol.~\bibinfo
  {volume} {22}\ (\bibinfo  {publisher} {Asociaci\'on F\'isica Argentina},\
  \bibinfo {year} {2010})\ pp.\ \bibinfo {pages} {37--41}\BibitemShut {NoStop}%
\bibitem [{\citenamefont {Moch\'an}\ \emph {et~al.}(2016)\citenamefont
  {Moch\'an}, \citenamefont {Ortiz}, \citenamefont {Mendoza},\ and\
  \citenamefont {P\'erez-Huerta}}]{Mochan(2016)}%
  \BibitemOpen
  \bibfield  {author} {\bibinfo {author} {\bibfnamefont {W.~L.}\ \bibnamefont
  {Moch\'an}}, \bibinfo {author} {\bibfnamefont {G.}~\bibnamefont {Ortiz}},
  \bibinfo {author} {\bibfnamefont {B.~S.}\ \bibnamefont {Mendoza}}, \ and\
  \bibinfo {author} {\bibfnamefont {J.~S.}\ \bibnamefont {P\'erez-Huerta}},\
  }\href {https://metacpan.org/pod/Photonic} {\enquote {\bibinfo {title}
  {Photonic},}\ }\bibinfo {howpublished} {Comprehensive Perl Archive Network
  (CPAN)} (\bibinfo {year} {2016}),\ \bibinfo {note} {perl package for
  calculations on metamaterials and photonic structures}\BibitemShut {NoStop}%
\bibitem [{\citenamefont {Ortiz}\ \emph {et~al.}(2014)\citenamefont {Ortiz},
  \citenamefont {Inchaussandague}, \citenamefont {Skigin}, \citenamefont
  {Depine},\ and\ \citenamefont {Moch\'an}}]{Ortiz(2014)}%
  \BibitemOpen
  \bibfield  {author} {\bibinfo {author} {\bibfnamefont {G.}~\bibnamefont
  {Ortiz}}, \bibinfo {author} {\bibfnamefont {M.}~\bibnamefont
  {Inchaussandague}}, \bibinfo {author} {\bibfnamefont {D.}~\bibnamefont
  {Skigin}}, \bibinfo {author} {\bibfnamefont {R.}~\bibnamefont {Depine}}, \
  and\ \bibinfo {author} {\bibfnamefont {W.~L.}\ \bibnamefont {Moch\'an}},\
  }\href@noop {} {\bibfield  {journal} {\bibinfo  {journal} {Journal of
  Optics}\ }\textbf {\bibinfo {volume} {16}},\ \bibinfo {pages} {105012}
  (\bibinfo {year} {2014})}\BibitemShut {NoStop}%
\bibitem [{\citenamefont {AG~Cullis}\ and\ \citenamefont
  {Calcott}(1997)}]{Canham(1997)}%
  \BibitemOpen
  \bibfield  {author} {\bibinfo {author} {\bibfnamefont {L.~C.}\ \bibnamefont
  {AG~Cullis}}\ and\ \bibinfo {author} {\bibfnamefont {P.}~\bibnamefont
  {Calcott}},\ }\href@noop {} {\bibfield  {journal} {\bibinfo  {journal} {J App
  Phys}\ }\textbf {\bibinfo {volume} {82}},\ \bibinfo {pages} {909} (\bibinfo
  {year} {1997})}\BibitemShut {NoStop}%
\bibitem [{\citenamefont {Bessa\"{\i}s}\ \emph {et~al.}(1996)\citenamefont
  {Bessa\"{\i}s}, \citenamefont {Ezzaouia}, \citenamefont {Elhouichet},
  \citenamefont {Oueslati},\ and\ \citenamefont {Bennaceur}}]{Bessais(1996)}%
  \BibitemOpen
  \bibfield  {author} {\bibinfo {author} {\bibfnamefont {B.}~\bibnamefont
  {Bessa\"{\i}s}}, \bibinfo {author} {\bibfnamefont {H.}~\bibnamefont
  {Ezzaouia}}, \bibinfo {author} {\bibfnamefont {H.}~\bibnamefont
  {Elhouichet}}, \bibinfo {author} {\bibfnamefont {M.}~\bibnamefont
  {Oueslati}}, \ and\ \bibinfo {author} {\bibfnamefont {R.}~\bibnamefont
  {Bennaceur}},\ }\href@noop {} {\bibfield  {journal} {\bibinfo  {journal}
  {Semicond. Sci. Technol.}\ }\textbf {\bibinfo {volume} {11}},\ \bibinfo
  {pages} {1815–1820} (\bibinfo {year} {1996})}\BibitemShut {NoStop}%
\bibitem [{\citenamefont {Shi}\ \emph {et~al.}(1993)\citenamefont {Shi},
  \citenamefont {Zheng}, \citenamefont {Wang},\ and\ \citenamefont
  {Yuan}}]{Shi(1993)}%
  \BibitemOpen
  \bibfield  {author} {\bibinfo {author} {\bibfnamefont {H.}~\bibnamefont
  {Shi}}, \bibinfo {author} {\bibfnamefont {Y.}~\bibnamefont {Zheng}}, \bibinfo
  {author} {\bibfnamefont {Y.}~\bibnamefont {Wang}}, \ and\ \bibinfo {author}
  {\bibfnamefont {R.}~\bibnamefont {Yuan}},\ }\href@noop {} {\bibfield
  {journal} {\bibinfo  {journal} {Appl. Phys. Lett.}\ }\textbf {\bibinfo
  {volume} {63}},\ \bibinfo {pages} {770} (\bibinfo {year} {1993})}\BibitemShut
  {NoStop}%
\bibitem [{\citenamefont {Billat}\ \emph {et~al.}(1995)\citenamefont {Billat},
  \citenamefont {Gaspard}, \citenamefont {Hérino}, \citenamefont {Ligeon},
  \citenamefont {Muller}, \citenamefont {Romestain},\ and\ \citenamefont
  {Vial}}]{Billat(1995)}%
  \BibitemOpen
  \bibfield  {author} {\bibinfo {author} {\bibfnamefont {S.}~\bibnamefont
  {Billat}}, \bibinfo {author} {\bibfnamefont {F.}~\bibnamefont {Gaspard}},
  \bibinfo {author} {\bibfnamefont {R.}~\bibnamefont {Hérino}}, \bibinfo
  {author} {\bibfnamefont {M.}~\bibnamefont {Ligeon}}, \bibinfo {author}
  {\bibfnamefont {F.}~\bibnamefont {Muller}}, \bibinfo {author} {\bibfnamefont
  {F.}~\bibnamefont {Romestain}}, \ and\ \bibinfo {author} {\bibfnamefont
  {J.}~\bibnamefont {Vial}},\ }\href@noop {} {\bibfield  {journal} {\bibinfo
  {journal} {Thin Solid Films}\ }\textbf {\bibinfo {volume} {263}},\ \bibinfo
  {pages} {238} (\bibinfo {year} {1995})}\BibitemShut {NoStop}%
\bibitem [{\citenamefont {Koshida}\ and\ \citenamefont
  {Koyama}(1992)}]{Koshida(1992)}%
  \BibitemOpen
  \bibfield  {author} {\bibinfo {author} {\bibfnamefont {N.}~\bibnamefont
  {Koshida}}\ and\ \bibinfo {author} {\bibfnamefont {H.}~\bibnamefont
  {Koyama}},\ }\href@noop {} {\bibfield  {journal} {\bibinfo  {journal} {Appl.
  Phys. Lett}\ }\textbf {\bibinfo {volume} {60}},\ \bibinfo {pages} {347}
  (\bibinfo {year} {1992})}\BibitemShut {NoStop}%
\bibitem [{\citenamefont {R.}\ \emph {et~al.}(2002)\citenamefont {R.},
  \citenamefont {R.Peña-Sierra},\ and\ \citenamefont
  {Castillo-Cabrera}}]{Paredes(2002)}%
  \BibitemOpen
  \bibfield  {author} {\bibinfo {author} {\bibfnamefont {G.}~\bibnamefont
  {R.}}, \bibinfo {author} {\bibnamefont {R.Peña-Sierra}}, \ and\ \bibinfo
  {author} {\bibfnamefont {G.}~\bibnamefont {Castillo-Cabrera}},\ }\href@noop
  {} {\bibfield  {journal} {\bibinfo  {journal} {Rev. Mex. Fis.}\ }\textbf
  {\bibinfo {volume} {48}},\ \bibinfo {pages} {92} (\bibinfo {year}
  {2002})}\BibitemShut {NoStop}%
\bibitem [{\citenamefont {Steiner}\ \emph {et~al.}(1996)\citenamefont
  {Steiner}, \citenamefont {Wiedenhofer}, \citenamefont {Kozlowski},\ and\
  \citenamefont {Lang}}]{Steiner(1996)}%
  \BibitemOpen
  \bibfield  {author} {\bibinfo {author} {\bibfnamefont {P.}~\bibnamefont
  {Steiner}}, \bibinfo {author} {\bibfnamefont {A.}~\bibnamefont
  {Wiedenhofer}}, \bibinfo {author} {\bibfnamefont {F.}~\bibnamefont
  {Kozlowski}}, \ and\ \bibinfo {author} {\bibfnamefont {W.}~\bibnamefont
  {Lang}},\ }\href@noop {} {\bibfield  {journal} {\bibinfo  {journal} {Thin
  Solid Films}\ }\textbf {\bibinfo {volume} {276}},\ \bibinfo {pages} {159}
  (\bibinfo {year} {1996})}\BibitemShut {NoStop}%
\bibitem [{\citenamefont {Martin}(2010)}]{Martin(2010)}%
  \BibitemOpen
  \bibfield  {author} {\bibinfo {author} {\bibfnamefont {P.~M.}\ \bibnamefont
  {Martin}},\ }\href@noop {} {\emph {\bibinfo {title} {Handbook of Deposition
  Technologies for Films and Coatings: Science, Applications and Technology}}}\
  (\bibinfo  {publisher} {Elsevier, USA},\ \bibinfo {year} {2010})\ p.\
  \bibinfo {pages} {740}\BibitemShut {NoStop}%
\bibitem [{\citenamefont {Bunshah}(1994)}]{Bunshah(1994)}%
  \BibitemOpen
  \bibfield  {author} {\bibinfo {author} {\bibfnamefont {R.~F.}\ \bibnamefont
  {Bunshah}},\ }\href@noop {} {\emph {\bibinfo {title} {Handbook of Deposition
  Technologies for Films and Coatings: Science, Applications and Technology}}}\
  (\bibinfo  {publisher} {Noyes Publications, USA},\ \bibinfo {year}
  {1994})\BibitemShut {NoStop}%
\bibitem [{\citenamefont {Cohen}(1983)}]{Cohen(1983)}%
  \BibitemOpen
  \bibfield  {author} {\bibinfo {author} {\bibfnamefont {S.~S.}\ \bibnamefont
  {Cohen}},\ }\href@noop {} {\bibfield  {journal} {\bibinfo  {journal} {Thin
  Solid Films}\ }\textbf {\bibinfo {volume} {104}},\ \bibinfo {pages} {361}
  (\bibinfo {year} {1983})}\BibitemShut {NoStop}%
\bibitem [{\citenamefont {Zerbino}\ \emph {et~al.}(2007)\citenamefont
  {Zerbino}, \citenamefont {Pesetti},\ and\ \citenamefont
  {Sustersic}}]{Zerbino(2007)}%
  \BibitemOpen
  \bibfield  {author} {\bibinfo {author} {\bibfnamefont {J.}~\bibnamefont
  {Zerbino}}, \bibinfo {author} {\bibfnamefont {L.}~\bibnamefont {Pesetti}}, \
  and\ \bibinfo {author} {\bibfnamefont {M.}~\bibnamefont {Sustersic}},\
  }\href@noop {} {\bibfield  {journal} {\bibinfo  {journal} {J. Mol. Liquids}\
  }\textbf {\bibinfo {volume} {131-132}},\ \bibinfo {pages} {185} (\bibinfo
  {year} {2007})}\BibitemShut {NoStop}%
\bibitem [{\citenamefont {Bishop}(2007)}]{Bishop(2007)}%
  \BibitemOpen
  \bibfield  {author} {\bibinfo {author} {\bibfnamefont {C.~A.}\ \bibnamefont
  {Bishop}},\ }\href@noop {} {\emph {\bibinfo {title} {Vacuum Deposition Onto
  Webs, Films, and Foils}}}\ (\bibinfo  {publisher} {Willian Andrew, New
  York},\ \bibinfo {year} {2007})\BibitemShut {NoStop}%
\bibitem [{\citenamefont {Ortiz}\ \emph {et~al.}(2010)\citenamefont {Ortiz},
  \citenamefont {Valdez}, \citenamefont {L\'opez}, \citenamefont {Mendoza},\
  and\ \citenamefont {Moch\'an}}]{Ortiz(2011)}%
  \BibitemOpen
  \bibfield  {author} {\bibinfo {author} {\bibfnamefont {G.}~\bibnamefont
  {Ortiz}}, \bibinfo {author} {\bibfnamefont {L.}~\bibnamefont {Valdez}},
  \bibinfo {author} {\bibfnamefont {G.}~\bibnamefont {L\'opez}}, \bibinfo
  {author} {\bibfnamefont {B.}~\bibnamefont {Mendoza}}, \ and\ \bibinfo
  {author} {\bibfnamefont {W.}~\bibnamefont {Moch\'an}},\ }\href@noop {}
  {\bibfield  {journal} {\bibinfo  {journal} {ANALES AFA}\ }\textbf {\bibinfo
  {volume} {22}} (\bibinfo {year} {2010})}\BibitemShut {NoStop}%
\bibitem [{\citenamefont {Johnson}\ and\ \citenamefont
  {Christy}(1972)}]{Johnson(1972)}%
  \BibitemOpen
  \bibfield  {author} {\bibinfo {author} {\bibfnamefont {P.~B.}\ \bibnamefont
  {Johnson}}\ and\ \bibinfo {author} {\bibfnamefont {R.~M.}\ \bibnamefont
  {Christy}},\ }\href@noop {} {\bibfield  {journal} {\bibinfo  {journal} {Phys.
  Rev. B}\ }\textbf {\bibinfo {volume} {6}},\ \bibinfo {pages} {4370} (\bibinfo
  {year} {1972})}\BibitemShut {NoStop}%
\bibitem [{\citenamefont {E.D.Palik}(1985)}]{Palik(1985)}%
  \BibitemOpen
  \bibinfo {editor} {\bibnamefont {E.D.Palik}},\ ed.,\ \href@noop {} {\emph
  {\bibinfo {title} {Handbook of optical constants of solids}}},\ Academic
  press handbook series\ (\bibinfo  {publisher} {Academic, Orlando, Florida.},\
  \bibinfo {year} {1985})\BibitemShut {NoStop}%
\bibitem [{\citenamefont {Garland}\ and\ \citenamefont
  {Tanner}(1978)}]{Etopim(1977)}%
  \BibitemOpen
  \bibinfo {editor} {\bibfnamefont {J.~C.}\ \bibnamefont {Garland}}\ and\
  \bibinfo {editor} {\bibfnamefont {D.~B.}\ \bibnamefont {Tanner}},\ eds.,\
  \href@noop {} {\emph {\bibinfo {title} {AIP Conference Proceeding}}},\ AIP
  Conference Proceeding No. 40\ (\bibinfo  {publisher} {American Institute of
  Physics, New York},\ \bibinfo {year} {1978})\BibitemShut {NoStop}%
\bibitem [{\citenamefont {Quintanilla}\ \emph {et~al.}(2000)\citenamefont
  {Quintanilla}, \citenamefont {Torquato},\ and\ \citenamefont
  {Ziff}}]{Quintanilla(2000)}%
  \BibitemOpen
  \bibfield  {author} {\bibinfo {author} {\bibfnamefont {J.}~\bibnamefont
  {Quintanilla}}, \bibinfo {author} {\bibfnamefont {S.}~\bibnamefont
  {Torquato}}, \ and\ \bibinfo {author} {\bibfnamefont {R.~M.}\ \bibnamefont
  {Ziff}},\ }\href {\doibase 10.1088/0305-4470/33/42/10} {\bibfield  {journal}
  {\bibinfo  {journal} {J. Phys. A: Math. Gen}\ }\textbf {\bibinfo {volume}
  {33}},\ \bibinfo {pages} {L399} (\bibinfo {year} {2000})}\BibitemShut
  {NoStop}%
\bibitem [{\citenamefont {Mochán}\ \emph {et~al.}(2015)\citenamefont
  {Mochán}, \citenamefont {Mendoza},\ and\ \citenamefont
  {Ortiz}}]{Mochan(2015)}%
  \BibitemOpen
  \bibfield  {author} {\bibinfo {author} {\bibfnamefont {W.~L.}\ \bibnamefont
  {Mochán}}, \bibinfo {author} {\bibfnamefont {B.~S.}\ \bibnamefont
  {Mendoza}}, \ and\ \bibinfo {author} {\bibfnamefont {G.~P.}\ \bibnamefont
  {Ortiz}},\ }in\ \href@noop {} {\emph {\bibinfo {booktitle} {Memorias del VI
  Taller sobre Metamateriales, Cristales Fotónicos, Cristales Fonónicos y
  Estructuras Plasmónicas}}}\ (\bibinfo  {publisher} {UNISON, Hermosillo, feb.
  2016},\ \bibinfo {address} {San Miguel de Allende},\ \bibinfo {year} {2015})\
  pp.\ \bibinfo {pages} {12--14}\BibitemShut {NoStop}%
\bibitem [{\citenamefont {Shalaev}(1996)}]{Shalaev(1996)}%
  \BibitemOpen
  \bibfield  {author} {\bibinfo {author} {\bibfnamefont {V.}~\bibnamefont
  {Shalaev}},\ }\href@noop {} {\bibfield  {journal} {\bibinfo  {journal}
  {Physics Reports}\ }\textbf {\bibinfo {volume} {272}},\ \bibinfo {pages} {61}
  (\bibinfo {year} {1996})}\BibitemShut {NoStop}%
\bibitem [{\citenamefont {Markel}\ \emph {et~al.}(1999)\citenamefont {Markel},
  \citenamefont {Shalaev}, \citenamefont {Zhang}, \citenamefont {Huynh},
  \citenamefont {L.Tay}, \citenamefont {Haslett},\ and\ \citenamefont
  {Moskovits}}]{Markel(1999)}%
  \BibitemOpen
  \bibfield  {author} {\bibinfo {author} {\bibfnamefont {V.}~\bibnamefont
  {Markel}}, \bibinfo {author} {\bibfnamefont {V.}~\bibnamefont {Shalaev}},
  \bibinfo {author} {\bibfnamefont {P.}~\bibnamefont {Zhang}}, \bibinfo
  {author} {\bibfnamefont {W.}~\bibnamefont {Huynh}}, \bibinfo {author}
  {\bibnamefont {L.Tay}}, \bibinfo {author} {\bibfnamefont {T.}~\bibnamefont
  {Haslett}}, \ and\ \bibinfo {author} {\bibfnamefont {M.}~\bibnamefont
  {Moskovits}},\ }\href@noop {} {\bibfield  {journal} {\bibinfo  {journal}
  {Phys. Rev. B.}\ }\textbf {\bibinfo {volume} {59}},\ \bibinfo {pages} {10903}
  (\bibinfo {year} {1999})}\BibitemShut {NoStop}%
\bibitem [{\citenamefont {M.I.Stockman}\ \emph {et~al.}(1996)\citenamefont
  {M.I.Stockman}, \citenamefont {L.N.Pandey},\ and\ \citenamefont
  {T.F.George}}]{Stockman(1996)}%
  \BibitemOpen
  \bibfield  {author} {\bibinfo {author} {\bibnamefont {M.I.Stockman}},
  \bibinfo {author} {\bibnamefont {L.N.Pandey}}, \ and\ \bibinfo {author}
  {\bibnamefont {T.F.George}},\ }\href@noop {} {\bibfield  {journal} {\bibinfo
  {journal} {Phys. Rev. B}\ }\textbf {\bibinfo {volume} {53}},\ \bibinfo
  {pages} {2183} (\bibinfo {year} {1996})}\BibitemShut {NoStop}%
\bibitem [{\citenamefont {Ortiz}\ and\ \citenamefont
  {Moch\'an}(2003)}]{Ortiz(2001)}%
  \BibitemOpen
  \bibfield  {author} {\bibinfo {author} {\bibfnamefont {G.}~\bibnamefont
  {Ortiz}}\ and\ \bibinfo {author} {\bibfnamefont {W.}~\bibnamefont
  {Moch\'an}},\ }\href@noop {} {\bibfield  {journal} {\bibinfo  {journal}
  {Phys. Rev. B.}\ }\textbf {\bibinfo {volume} {67}} (\bibinfo {year}
  {2003})}\BibitemShut {NoStop}%
\bibitem [{\citenamefont {Moskovits}(1985)}]{Moskovits(1985)}%
  \BibitemOpen
  \bibfield  {author} {\bibinfo {author} {\bibfnamefont {M.}~\bibnamefont
  {Moskovits}},\ }\href@noop {} {\bibfield  {journal} {\bibinfo  {journal}
  {Rev. Mod. Phys.}\ }\textbf {\bibinfo {volume} {5}},\ \bibinfo {pages} {783}
  (\bibinfo {year} {1985})}\BibitemShut {NoStop}%
\bibitem [{\citenamefont {Sarychev}\ and\ \citenamefont
  {Shalaev}(2000)}]{Sarychev(2000)}%
  \BibitemOpen
  \bibfield  {author} {\bibinfo {author} {\bibfnamefont {A.}~\bibnamefont
  {Sarychev}}\ and\ \bibinfo {author} {\bibfnamefont {V.}~\bibnamefont
  {Shalaev}},\ }\href@noop {} {\bibfield  {journal} {\bibinfo  {journal}
  {Physics Reports}\ }\textbf {\bibinfo {volume} {335}},\ \bibinfo {pages}
  {275} (\bibinfo {year} {2000})}\BibitemShut {NoStop}%
\bibitem [{\citenamefont {Ossicini}\ \emph {et~al.}(2003)\citenamefont
  {Ossicini}, \citenamefont {Pavesi},\ and\ \citenamefont
  {Priolo}}]{Pavesi(2003)}%
  \BibitemOpen
  \bibfield  {author} {\bibinfo {author} {\bibfnamefont {S.}~\bibnamefont
  {Ossicini}}, \bibinfo {author} {\bibfnamefont {L.}~\bibnamefont {Pavesi}}, \
  and\ \bibinfo {author} {\bibfnamefont {F.}~\bibnamefont {Priolo}},\
  }\href@noop {} {\emph {\bibinfo {title} {Light Emiting Silicon For
  Microphotonics}}},\ Vol.\ \bibinfo {volume} {194}\ (\bibinfo  {publisher}
  {Springer},\ \bibinfo {year} {2003})\BibitemShut {NoStop}%
\bibitem [{\citenamefont {Toranzos}\ \emph {et~al.}(2008)\citenamefont
  {Toranzos}, \citenamefont {Koropecki}, \citenamefont {Urteaga},\ and\
  \citenamefont {Ortiz}}]{Toranzos(2008)b}%
  \BibitemOpen
  \bibfield  {author} {\bibinfo {author} {\bibfnamefont {V.}~\bibnamefont
  {Toranzos}}, \bibinfo {author} {\bibfnamefont {R.}~\bibnamefont {Koropecki}},
  \bibinfo {author} {\bibfnamefont {R.}~\bibnamefont {Urteaga}}, \ and\
  \bibinfo {author} {\bibfnamefont {G.}~\bibnamefont {Ortiz}},\ }\href@noop {}
  {\bibfield  {journal} {\bibinfo  {journal} {ANALES AFA}\ }\textbf {\bibinfo
  {volume} {20}},\ \bibinfo {pages} {115} (\bibinfo {year} {2008})}\BibitemShut
  {NoStop}%
\bibitem [{\citenamefont {Buda}\ \emph {et~al.}(1992)\citenamefont {Buda},
  \citenamefont {Kohanoff},\ and\ \citenamefont {Parrinello}}]{Buda(1992)}%
  \BibitemOpen
  \bibfield  {author} {\bibinfo {author} {\bibfnamefont {F.}~\bibnamefont
  {Buda}}, \bibinfo {author} {\bibfnamefont {J.}~\bibnamefont {Kohanoff}}, \
  and\ \bibinfo {author} {\bibfnamefont {M.}~\bibnamefont {Parrinello}},\
  }\href@noop {} {\bibfield  {journal} {\bibinfo  {journal} {Phys. Rev. Lett.}\
  }\textbf {\bibinfo {volume} {69}},\ \bibinfo {pages} {1272} (\bibinfo {year}
  {1992})}\BibitemShut {NoStop}%
\bibitem [{\citenamefont {Kovalev}\ \emph {et~al.}(1999)\citenamefont
  {Kovalev}, \citenamefont {Heckler}, \citenamefont {Polisski},\ and\
  \citenamefont {Koch}}]{Kovalev(1999)}%
  \BibitemOpen
  \bibfield  {author} {\bibinfo {author} {\bibfnamefont {D.}~\bibnamefont
  {Kovalev}}, \bibinfo {author} {\bibfnamefont {H.}~\bibnamefont {Heckler}},
  \bibinfo {author} {\bibfnamefont {G.}~\bibnamefont {Polisski}}, \ and\
  \bibinfo {author} {\bibfnamefont {F.}~\bibnamefont {Koch}},\ }\href@noop {}
  {\bibfield  {journal} {\bibinfo  {journal} {Phys. Stat. Sol. B}\ }\textbf
  {\bibinfo {volume} {215}},\ \bibinfo {pages} {871} (\bibinfo {year}
  {1999})}\BibitemShut {NoStop}%
\bibitem [{\citenamefont {Fauchet}\ and\ \citenamefont {von
  Behren}(1997)}]{Fauchet(1997)}%
  \BibitemOpen
  \bibfield  {author} {\bibinfo {author} {\bibfnamefont {P.}~\bibnamefont
  {Fauchet}}\ and\ \bibinfo {author} {\bibfnamefont {J.}~\bibnamefont {von
  Behren}},\ }\href@noop {} {\bibfield  {journal} {\bibinfo  {journal} {Phys.
  Stat. Sol. B}\ }\textbf {\bibinfo {volume} {204}},\ \bibinfo {pages} {R7}
  (\bibinfo {year} {1997})}\BibitemShut {NoStop}%
\bibitem [{\citenamefont {Bohren}\ and\ \citenamefont
  {Huffman}(1983)}]{Bohren(1983)}%
  \BibitemOpen
  \bibfield  {author} {\bibinfo {author} {\bibfnamefont {C.~F.}\ \bibnamefont
  {Bohren}}\ and\ \bibinfo {author} {\bibfnamefont {D.~R.}\ \bibnamefont
  {Huffman}},\ }\href@noop {} {\emph {\bibinfo {title} {Absorption and
  Scattering of light by small particles}}}\ (\bibinfo  {publisher} {John Wiley
  and Sons, Inc.},\ \bibinfo {year} {1983})\BibitemShut {NoStop}%
\end{thebibliography}%

\end{document}